\documentclass[12pt]{article}
\usepackage{amsmath}

\usepackage{graphicx}
\usepackage{comment}
\renewcommand{\d}{{\rm d}}
\newcommand{\bx}{\mbox{\boldmath$x$}}
\newcommand{\by}{\mbox{\boldmath$y$}}
\newcommand{\bn}{\mbox{\boldmath$n$}}
\newcommand{\bN}{\mbox{\boldmath$N$}}
\newcommand{\bDelta}{\overline{\Delta}}
\newcommand{\bp}{\mbox{\boldmath$p$}}
\newcommand{\rl}{\ell} 
\newcommand{\bnabla}{\mbox{\boldmath$\nabla$}}
\newcommand{\ba}{\mbox{\boldmath$a$}}
\newcommand{\bl}{\mbox{\boldmath$l$}}
\newcommand{\bk}{\mbox{\boldmath$k$}}
\renewcommand{\Re}{\mathop{\rm Re}\nolimits}
\renewcommand{\Im}{\mathop{\rm Im}\nolimits}
\newcommand{\sgn}{\mathop{\rm sgn}\nolimits}
\newcommand{\tg}{\mathop{\rm tg}\nolimits}
\newcommand{\sh}{\mathop{\rm sh}\nolimits}
\newcommand{\ch}{\mathop{\rm ch}\nolimits}
\newcommand{\arsh}{\mathop{\rm arsh}\nolimits}
\newcommand{\tilS}{\tilde{S}}
\begin{document}

\title{On the discrete version of the Kerr geometry
}

\author{V.M. Khatsymovsky \\
 {\em Budker Institute of Nuclear Physics} \\ {\em of Siberian Branch Russian Academy of Sciences} \\ {\em
 Novosibirsk,
 630090,
 Russia}
\\ {\em E-mail address: khatsym@gmail.com}}
\date{}
\maketitle
\begin{abstract}
A Kerr type solution in the Regge calculus is considered. It is assumed that the discrete general relativity, the Regge calculus, is quantized within the path integral approach. The only consequence of this approach used here is the existence of a length scale at which edge lengths are loosely fixed, as considered in our earlier paper.

In addition, we previously considered the Regge action on a simplicial manifold on which the vertices are coordinatized and the corresponding piecewise constant metric introduced, and found that for the simplest periodic simplicial structure and in the leading order over metric variations between 4-simplices, this reduces to a finite-difference form of the Hilbert-Einstein action.

The problem of solving the corresponding discrete Einstein equations (classical) with a length scale (having a quantum nature) arises as the problem of determining the optimal background metric for the perturbative expansion generated by the functional integral. Using an one-complex-function ansatz for the metric, which reduces to the Kerr-Schild metric in the continuum, we find a discrete metric that approximates the continuum one at large distances and is nonsingular on the (earlier) singularity ring. The effective curvature $R_{\lambda \nu \nu \rho }$, including where $R_{\lambda \mu} \neq 0$ (gravity sources), is analyzed with a focus on the vicinity of the singularity ring.
\end{abstract}


MSC classes: 83C27; 83C57

keywords: Einstein theory of gravity; minisuperspace model; piecewise flat spacetime; Regge calculus; Kerr black hole

\section{Introduction}

General relativity (GR), formulated on a subclass of the Riemannian manifolds, the piecewise flat manifolds or simplicial complexes composed of flat 4-dimensional tetrahedra or 4-simplices, was proposed by Regge in 1961 \cite{Regge} as a discrete approximation to ordinary GR on smooth manifolds. The conviction that in this case we do not lose essential degrees of freedom is based on the possibility of using a piecewise flat manifold to arbitrarily accurately approximate an arbitrary Riemannian manifold \cite{Fein,CMS}. This formulation fits well into the discrete approach to defining the functional integral in quantum gravity \cite{Ham}, a proven tool for extracting physical quantities \cite{HamWil1,HamWil2}. Regge's discrete coordinateless formulation serves here as a regularization tool that is not tied to any coordinate system. The functional integral can be regularized in a similar, but more computable way, by limiting the set of 4-simplices to several of their types in the Causal Dynamical Triangulations approach \cite{cdt}. It can also be assumed that the structure of spacetime at small distances is given by a piecewise flat manifold \cite{Mik}.

The application of the Regge calculus to cosmology was considered in the works \cite{WilCol,Gen,Bre4,WilLiu}. There we are talking about the numerical analysis of cosmological models; their emergence in the originally quantum Causal Dynamical Triangulations approach was considered in \cite{GlaLol}. Approximation by a simplicial complex was probably first applied to the geometries of Schwarzschild and Reissner-Nordström in \cite{Wong}, where it was used for numerical estimates in GR. Subsequently, in \cite{Bre2}, for such systems, more efficient computational lattice methods in numerical GR were proposed. Quantum analysis in the framework of Loop Quantum Gravity was applied to the Schwarzschild black hole in \cite{Ash1,Ash2}. In this system, the singularity was resolved due to a nonzero area quantum (of the order of the Plank scale), which plays here the same regularizing role as the elementary area in lattice theory.

We have analyzed the Schwarzschild black hole \cite{Kha1}, including the case of slow rotation (Lense-Thirring metric) \cite{our} in simplicial gravity. The approach of these papers, which minimally involves the concept of a path integral, is reduced to solving classical discrete equations in which the elementary length, which plays the role of the lattice spacing, is of quantum origin. We now aim to analyze a discrete version of the Kerr black hole. We first briefly describe the approach in Section \ref{method}. Section \ref{general} gives a metric ansatz for which discrete equations are to be solved. The solution to these discrete equations can be obtained from a similar solution to the discrete Schwarzschild problem using analytic continuation, and continuation issues are discussed in Section \ref{continuation}. The discrete metric in a neighborhood of the singular points (more exactly, those that were singular in the continuum) is found in Section \ref{metric}. The Riemann tensor when approaching the singular points from the vacuum side (that is, from the points where the Ricci tensor is equal to zero) is found in Section \ref{riemann}. The distribution of gravity sources (that is, of nonzero Ricci or Einstein tensors), mainly in the vicinity of the singular points, is found in Section \ref{sources}.

\section{The method}\label{method}

We can think of triangulation as a measurement process that creates the corresponding simplicial complex and piecewise flat spacetime as the simplest structure that describes this triangulation (as opposed to the case where 4-simplices have intrinsic curvature). Then it is important that there is a mechanism that fixes the scale of edge lengths (or elementary lengths) at some nonzero level \cite{our1}. This is due to a specific form of the functional integral measure, which has a maximum at certain values of the edge lengths. Ideally, the Regge calculus strategy implies averaging over possible simplicial structures; consider here the case of some fixed structure.

To get a well-defined functional integral in Regge calculus, an appropriate choice of variables is important. In this respect, the connection variables introduced by Fr\"{o}hlich \cite{Fro} are promising. We take Regge action in terms of both tetrad variables (edge vectors) and independent on them connection matrices or SO(3,1) rotations living on the tetrahedra or 3-simplices, in particular, set the action to be a combination of contributions from self-dual and anti-self-dual connection matrices separately \cite{Kha}. The use of connection variables is motivated by difficulties of passing in a non-singular way to the continuous time limit and to the canonical Hamiltonian form of the action in terms of purely metric (length or vector) variables. Excluding the connection classically gives the original Regge action.

With such a form of the action in the larger configuration superspace of length/vec\-tor and connection variables, the Jacobian of the Poisson brackets of constraints, which determines the canonical path integral measure, still has singularities, which is typical for discrete theories, but we can define such a measure in a non-singular way in a more extended configuration superspace in which the area tensors living on the triangles or 2-simplices are independent. Then we can restore the full discrete measure which reduces to the found canonical one in the continuous time limit, no matter what coordinate is chosen as time. This full measure can be projected onto the physical hypersurface of unambiguously defined edge lengths in the extended configuration superspace by inserting the corresponding $\delta$-function factor. In the resulting functional integral, we can perform integration over the connection. After functional integration over the connection, we have a functional integral in terms of edge vectors only,
\begin{equation}                                                            
\int \exp [ i S(\rl , \Omega  ) ] ( \cdot ) \d \mu ( \rl ) {\cal D} \Omega = \int \exp [ i \tilS ( \rl ) ] ( \cdot ) F ( \rl ) D \rl ,
\end{equation}

\noindent as a functional on the functions of the set of the edge lengths $\rl = (l_1, \dots, l_n )$. Here we can use different methods to compute the modulus $F ( \rl )$ (measure) and the argument $\tilS ( \rl )$ (phase) of the result, which lead to expansions over different parameters. Namely, parameterizing metric by the ADM lapse-shift functions\cite{ADM1}, we extract $F ( \rl )$ from the expansion of the functional integral over the scale $N$ of the discrete lapse-shifts $(N, \bN )$ (certain edge vectors). $\tilS ( \rl )$ is extracted from the stationary phase expansion of the functional integral. The parameters of these expansions both can be small if the length scale $b$ (\ref{b=sqrt}) below is formally large in the Planck scale $\sqrt{G}$ units. Using namely such expansions for $F ( \rl )$ and $\tilS ( \rl )$ is justified by that nonzero contributions into $F ( \rl )$ and $\tilS ( \rl )$ arise just in the leading orders of the corresponding expansions of the functional integral.

In particular, $\tilS ( \rl )$ in the leading order of the stationary phase expansion is just the Regge action $S ( \rl )$, since it follows from $S(\rl , \Omega  )$ by excluding $\Omega$ classically by definition of $\tilS ( \rl )$ and construction of this expansion.

The measure in the path integral is such that the perturbative expansion generated by this integral provides a leading order contribution that has a maximum at a certain optimal initial point $\rl_0 = (l_{01}, \dots, l_{0n} )$ (background metric). Roughly speaking, this is the maximum point of the measure. In somewhat more detail, we pass from $\rl $ to some new variable $u = (u_1, \dots, u_n )$ that reduces the measure $F( \rl ) D \rl$ to the Lebesgue one $D u$,
\begin{equation}\label{Sdudu}                                               
S (\rl ) = \frac{1}{2} \sum_{j, k, l, m} \frac{\partial^2 S (\rl_0 )}{\partial l_j \partial l_l} \frac{\partial l_j (u_0 )}{\partial u_k} \frac{\partial l_l (u_0 )}{\partial u_m} \Delta u_k \Delta  u_m + \dots ,
\end{equation}

\noindent $\Delta u = u - u_0$, at
\begin{equation}\label{dS/dl=0}                                             
\frac{\partial S(l_0)}{\partial l_j} = 0 ,
\end{equation}

\noindent and the determinant of the second order form in the exponent should have a minimum at this point, or
\begin{equation}\label{def-l0}                                              
F (\rl_0 )^2 \det \left \| \frac{\partial^2 S (\rl_0 )}{\partial l_i \partial l_k} \right \|^{-1}
\end{equation}

\noindent should have a maximum. This, in particular, sets the scale of the edge lengths (of the spatial and diagonal edges, but not of the temporal ones, whose vectors, the discrete analogs of the lapse-shift functions, are fixed by hand). The equations of motion (\ref{dS/dl=0}) themselves do not fix such a variable. The function (\ref{def-l0}) has a bell-shaped form and a pronounced maximum when the typical area of the spatial and diagonal triangles is $b^2 / 2$, where the typical length scale is
\begin{equation}\label{b=sqrt}                                              
b = \sqrt{ 32 G ( \eta - 5) / 3 } .
\end{equation}

\noindent Here, $\eta$ is a fundamental parameter that characterizes the quantum extension of the theory and parameterizes the factors $V_{\sigma^4}^\eta$ in the path-integral measure, where $V_{\sigma^4}$ is the 4-volume of the 4-simplex $\sigma^4$.

Now the task is to find a Kerr-like solution as an optimal background metric for the perturbative expansion in the framework of the functional integral discussed above. That is, we should solve the Regge skeleton equations (\ref{dS/dl=0}) together with the requirement to maximize the measure (\ref{def-l0}), which fixes the length scale $b$ (\ref{b=sqrt}).

We have considered \cite{our2} a simplicial complex, in which certain coordinates are given to the vertices, so that the geometry can be described by a piecewise constant metric. The Regge action can be calculated in terms of this metric through intermediate definition of discrete Christoffel symbols and written in the form of an expansion over metric variations from 4-simplex to 4-simplex. In particular, we can take the simplest periodic simplicial complex, the cell of which is topologically a cube divided into $4! = 24$ 4-simplices with the help of the diagonals drawn from one of its vertices \cite{RocWil}. Then the leading order term in this expansion is a finite-difference form of the Hilbert-Einstein action,
\begin{eqnarray}\label{DM+MM}                                               
\sum_{\rm 4-cubes} {\cal K}^{\lambda \mu}_{~~~ \lambda \mu} \sqrt{g} , ~~~ {\cal K}^\lambda_{~ \, \mu \nu \rho} \! = \! \Delta_\nu M^\lambda_{\rho \mu} \! - \! \Delta_\rho M^\lambda_{\nu \mu} \! + \! M^\lambda_{\nu \sigma} M^\sigma_{\rho \mu} \! - \! M^\lambda_{\rho \sigma} M^\sigma_{\nu \mu} , \nonumber \\ \hspace{-10mm}
M^\lambda_{\mu \nu} = \frac{1}{2} g^{\lambda \rho} (\Delta_\nu g_{\mu \rho} + \Delta_\mu g_{\rho \nu} - \Delta_\rho g_{\mu \nu}), ~~~ \Delta_\lambda = 1 - \overline{T}_\lambda .
\end{eqnarray}

\noindent $T_\lambda$ is the shift operator along the coordinate $x^\lambda$ by 1.

When calculating the functional measure, we use the expansion over the discrete lapse-shift vectors as constant parameters. That is, we can introduce coordinates of the vertices so that the metric included in (\ref{DM+MM}) has $(N, \bN ) = {\rm const}$. A particular case is the metric in a synchronous frame, $(N, \bN ) = (1, {\bf 0})$. In the leading order over metric variations, our task is reduced to solving the finite-difference form of the Einstein equations in a synchronous frame.

Kerr geometry can be described by convenient metrics in one case or another \cite{Baines}, we start with the Boyer-Lindquist metric \cite{Boyer},
\begin{eqnarray}\label{ds2-Boyer}                                           
\d s^2 & = & \left ( - 1 + \frac{r_g r}{ \rho^2 } \right ) \d t^2 + \frac{ \rho^2 }{ \triangle } \d r^2 + \rho^2 \d \theta^2 + \left ( r^2 + a^2 + a^2 \frac{r_g r}{ \rho^2 } \sin^2 \theta \right ) \sin^2 \theta \d \varphi^2 \nonumber \\ & & - 2 a \frac{r_g r}{ \rho^2 } \sin^2 \theta \d \varphi \d t , \\ & & \mbox{ where } \rho^2 = r^2 + a^2 \cos^2 \theta , ~ \triangle = r^2 - r_g r + a^2 . \nonumber
\end{eqnarray}

\noindent For this geometry, we can introduce a synchronous frame similar to that in the case of the Lemaitre metric for the Schwarzschild geometry. Such a metric for the Kerr geometry under consideration can be obtained in a similar way, by binding the coordinates to a set of freely moving particles, and we can write for it an expression of the form\cite{Kha2},
\begin{eqnarray}
& & \d s^2 = - \d \tau^2 + \frac{r r_1}{\rho^2} \d r_1^2 - 2 a^2 \frac{ r \sin ( 2 \theta )}{ \rho^2 } I \sqrt{ r_1 } \d r_1 \d \theta + 2 a \sqrt{ r_g } \frac{r \sin^2 \theta}{ \rho^2 } \sqrt{ r_1 } \d r_1 \d \varphi_1 \nonumber \\ & & + \left [ \rho^2 + a^4 \frac{ r \sin^2 ( 2 \theta )}{ \rho^2 } I^2 \right ] \d \theta^2 - 2 a^3 \sqrt{ r_g } \frac{ r \sin ( 2 \theta ) \sin^2 \theta}{ \rho^2 } I \d \theta \d \varphi_1 \nonumber \\ & & + \left ( r^2 + a^2 + a^2 \frac{r_g r}{ \rho^2 } \sin^2 \theta \right ) \sin^2 \theta\d \varphi_1^2 , \mbox{ where } I = \int_r^\infty \frac{ \d r }{ \sqrt{ r \left ( r^2 + a^2 \right )} } .
\end{eqnarray}

\noindent Here $r$ is a function of $\tau$, $r_1$, $\theta$ by means of
\begin{equation}\label{tau=tau(r1,r,theta)}
\tau = \frac{ 2 }{ 3 } \frac{ r_1^{ 3 / 2 } - r^{ 3 / 2 } }{ \sqrt{ r_g } } + \int_r^\infty \left [ \frac{ r^2 + a^2 \cos^2 \theta }{ \sqrt{ r \left ( r^2 + a^2 \right )} } - \sqrt{ r } \right ] \frac{ \d r }{ \sqrt{ r_g } } .
\end{equation}

\noindent By excluding $r_1$ in favor of $r$, we obtain an analogue of the Painlev\'{e}-Gullstrand metric for the Schwarzschild geometry, the Doran metric\cite{Doran},
\begin{eqnarray}\label{Doran}
& & \d s^2 = \left ( - 1 + \frac{r_g r}{\rho^2} \right ) \d \tau^2 + \frac{\rho^2}{r^2 + a^2} \d r^2 + \rho^2 \d \theta^2 + 2 \sqrt{ \frac{ r_g r }{ r^2 + a^2 } } \d \tau \d r \nonumber \\ & & + 2 a \sqrt{ \frac{ r_g r }{ r^2 + a^2 } } \sin^2 \theta \d r \d \varphi_1 + 2 a \frac{ r_g r }{ \rho^2 } \sin^2 \theta \d \tau \d \varphi_1 \nonumber \\ & & + \left ( r^2 + a^2 + a^2 \frac{r_g r}{ \rho^2 } \sin^2 \theta \right ) \sin^2 \theta \d \varphi_1^2 .
\end{eqnarray}

As a typical example, consider a triangulation of the system or a possible form of the simplicial complex, if possible, almost everywhere of the simplest periodic structure, in a section along $r_1$, $\tau$ passing through the singularity $r = 0$, $\theta = \pi / 2$. This section consists of two regions ($\varphi_1 = {\rm const}$ in them): $\theta = \pi / 2$ and $r$ varies from $\infty$ to $0$ (the equatorial plane outside the singularity ring); $r = 0$ and $\theta$ changes from $\pi / 2$ to $0$ (the equatorial plane inside the singularity ring).

If $\tau = {\rm const}$, $\varphi_1 = {\rm const}$, $\theta = \pi / 2$ and $r$ is changing, then the invariant length along the coordinate line $\d r_1$ or $\d r$ is determined as
\begin{equation}
\d s^2 = \frac{\rho^2 }{r^2 + a^2 } \d r^2 = \frac{r^2 }{r^2 + a^2 } \d r^2 , ~ \d s \approx \frac{r \d r}{ a } \mbox{ (at $r \ll a$) } , ~ s \approx \frac{r^2}{ 2 a }
\end{equation}

\noindent (according to $\d s^2$ in terms of $\tau$, $r$, $\theta$, $\varphi_1$ (\ref{Doran})), and if $s$ covers $n$ typical spacings, $s = n b$, then the $n$-th vertex has $r \approx \sqrt{2 a b n }$. If the singularity corresponds to $r_1 = r_{1(0)}$ at $\tau = \tau_0$, then (\ref{tau=tau(r1,r,theta)}) gives that the $n$-th vertex from it at the same $\tau$ corresponds to $r_1 = r_{1(n)}$, and the world line $r_1 = r_{1(n)}$ intersects $r = 0$ at $\tau = \tau_0 + \Delta \tau_n$ such that
\begin{equation}\label{r1(n)}
\frac{3}{2} \sqrt{ r_g } \Delta \tau_n = r_{1(n)}^{3 / 2} - r_{1(0)}^{3 / 2} = \frac{3}{2} \int_0^r \frac{r^{3 / 2}}{\sqrt{r^2 + a^2}} \d r \approx \frac{3}{5 a} r^{5 / 2} \approx \frac{3}{5 a} ( 2 a b n )^{5 / 4} .
\end{equation}

If $\tau = {\rm const}$, $\varphi_1 = {\rm const}$, $r = 0$ and $\theta$ is changing, the invariant length along the coordinate line $\d r_1$ or $\d \theta$ is determined as
\begin{equation}
\d s^2 = \rho^2 \d \theta^2 = a^2 \cos^2 \theta \d \theta^2 , s = a ( 1 - \sin \theta ) \approx \frac{a}{ 2 } \left ( \frac{ \pi }{ 2 } - \theta \right )^2 \mbox{ (at $\frac{2}{ \pi } \left | \frac{ \pi }{ 2 } - \theta \right | \ll 1$) }
\end{equation}

\noindent (according to $\d s^2$ in terms of $\tau$, $r$, $\theta$, $\varphi_1$ (\ref{Doran})), and if $s$ covers $n$ typical spacings, $s = n b$, then the $(-n)$-th vertex from $\theta = \pi / 2$ has $(\pi / 2 - \theta )^2 \approx 2 b n / a $. Then (\ref{tau=tau(r1,r,theta)}) gives that the $(-n)$-th vertex from the singularity at $\tau = \tau_0$ corresponds to $r_1 = r_{1(-n)}$, and the world line $r_1 = r_{1(-n)}$ intersects $r = 0$ at $\tau = \tau_0 + \Delta \tau_{-n}$ such that
\begin{eqnarray}\label{r1(-n)}
\frac{3}{2} \sqrt{ r_g } \Delta \tau_{-n} & = & r_{1(-n)}^{3 / 2} - r_{1(0)}^{3 / 2} = - \frac{3}{2} \int_0^\infty{ \frac{a^2 \cos^2 \theta}{\sqrt{r \left ( r^2 + a^2 \right )}} \d r} \approx - 6 I_0 a^{1 / 2} b n , \nonumber \\ I_0 & = & \int_0^\infty{ \frac{\d y}{ \sqrt{ 1 + y^4 } }}= [ \Gamma ( 1 / 4 ) ]^2 / ( 4 \sqrt{ \pi } ) = 1.854... .
\end{eqnarray}

\begin{figure}[ht]
\centerline{\includegraphics[width=12.5cm]{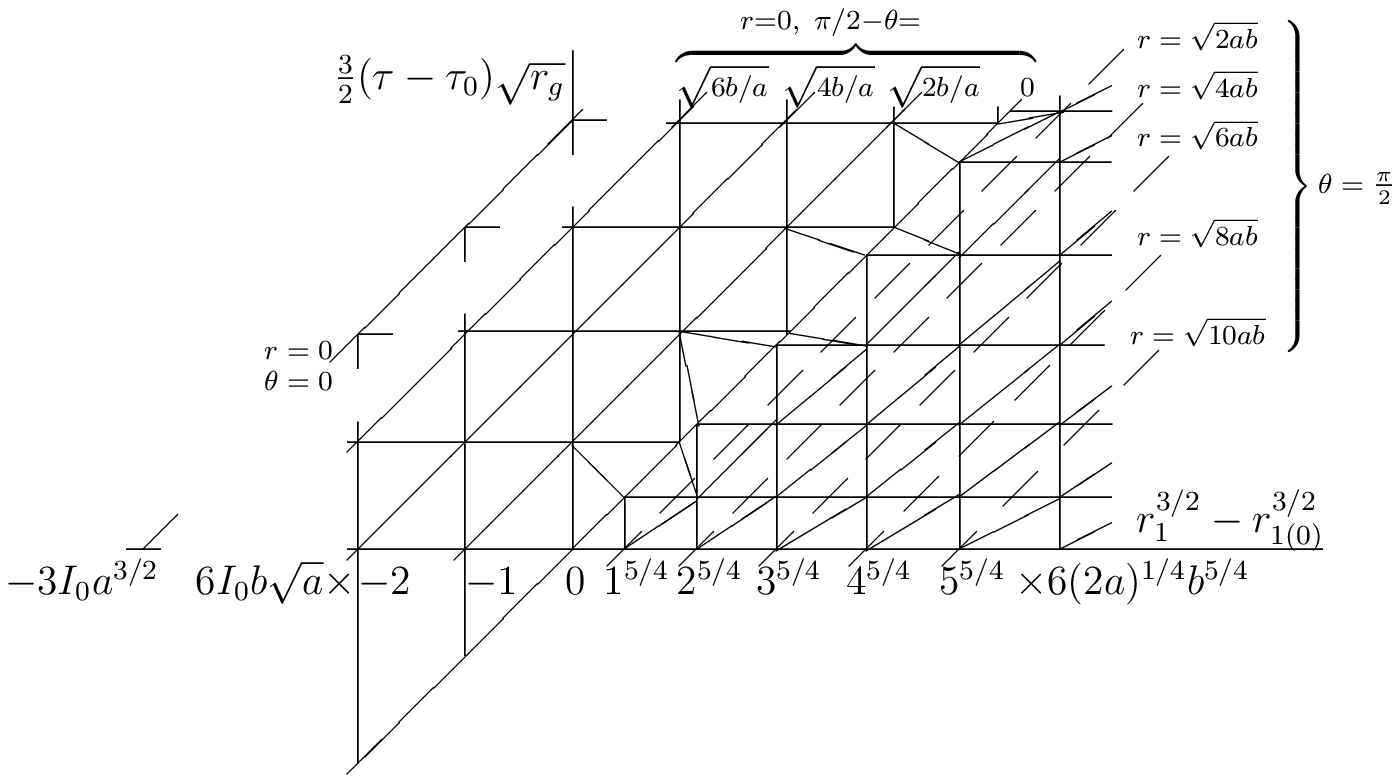}}
\caption{A triangulation using, if possible, almost everywhere the simplest periodic structure obtained from the cubic subdivision in the Lemaitre type coordinates in a section along $r_1$, $\tau$ passing through the singularity. \label{f1}}
\end{figure}

This is illustrated by Fig. \ref{f1}. The difference in $r_1$ between two neighboring world lines (\ref{r1(n)}) or (\ref{r1(-n)}), which in a sense characterizes the density of world lines, is different on both sides of the singularity, (\ref{r1(n)}) and (\ref{r1(-n)}). This leads to irregularities in the nearest neighborhood of the singularity in the simplest used periodic simplicial structure.

Of course, there is also a violation of the triangulation regularity connected with that we aim to choose $\Delta \tau \simeq b$ for the difference in $\tau$ between the leaves $\tau = {\rm const}$ to deal with 4-simplices that are not too shrunk in any direction. Normally the leaves go through the vertices at the intersections of the world lines $r_1 = {\rm const}$ with the singularity. But if the difference in $\tau$ between neighboring such vertices (defined by the difference in $r_1$), $\Delta \tau_{n + 1} - \Delta \tau_n$ (\ref{r1(n)}) or $\Delta \tau_{- n} - \Delta \tau_{- n - 1}$ (\ref{r1(-n)}) is significantly less or greater than $b$, then either only some of the vertices at the intersections of the lines $r_1 = {\rm const}$ with the singularity belong to a leaf $\tau = {\rm const}$ (the rest of these vertices are connected by edges with the nearest vertices of the nearest leaves) or additional leaves between the two under consideration are introduced (and the vertices at their intersections with the singularity are connected by edges with the nearest vertices in the $\tau = {\rm const}$ leaves). In both cases, there is again a violation of the triangulation regularity in the nearest neighborhood of the singularity.

In our analysis, we should consider the general simplicial metric, which is close at large distances in a certain topology to the Kerr one, here in a synchronous frame of reference. We use the simplest periodic (almost everywhere) simplicial complex obtained by a triangulation like the above one, and the action in the leading order over metric variations from simplex to simplex is a finite-difference form of the Hilbert-Einstein action (\ref{DM+MM}), where the interpolating smooth metric $g_{\lambda \mu}$ is close in a certain topology to the Kerr one. In principle, subsequent orders over metric variations could be taken into account; the corresponding terms should be written out in (\ref{DM+MM}); in these orders, we should distinguish between the metrics in different 4-simplices inside the 4-cube.

In the leading order over metric variations, just considered in the present paper, the situation is simplified. Finite differences obey the same rules as the corresponding derivatives; in particular, in this order there is "diffeomorphism invariance": in a finite-difference expression, we can go from a metric close to the Kerr metric in a synchronous frame of reference to a metric close to the Kerr metric in some other frame of reference. As the latter reference system, we will choose the Kerr-Schild coordinate system \cite{Kerr,Kerr1}. Now we do not need to know the details of triangulation in the intermediate synchronous frame; only its existence is important in principle.

\section{The metric ansatz}\label{general}

So, consider the original continuum metric in the Kerr-Schild coordinates,
\begin{eqnarray}\label{ds2Kerr}
\d s^2 & = & - \d \tau^2 + \d x_1^2 + \d x_2^2 + \d x_3^2 \nonumber \\ & & + \frac{ r_g r^3}{ r^4 + a^2 x_3^2} \left ( \d \tau - \frac{r x_1 - a x_2}{r^2 + a^2} \d x_1 - \frac{r x_2 + a x_1}{r^2 + a^2} \d x_2 - \frac{x_3 }{r } \d x_3 \right )^2 .
\end{eqnarray}

\noindent Here, for convenience, the coordinates $ x_k $ (whose differentials are in fact, of course, contravariant vectors) will be denoted by subscripts. The quantity $r$ obeys the equation connecting the radial oblate spheroidal coordinate $r$ with the Cartesian coordinates $x_1, x_2, x_3$,
\begin{equation}
r^4 - r^2 ( x_1^2 + x_2^2 + x_3^2 - a^2 ) - a^2 x_3^2 = 0 .
\end{equation}

This metric is a particular case of the algebraically special metric \cite{Kerr,Kerr1,Chandra}
\begin{equation}
g_{\lambda \mu} = \eta_{\lambda \mu} + l_\lambda l_\mu , ~~~ l^\lambda l_\lambda = 0 , ~~~ l^\lambda = \eta^{\lambda \mu} l_\mu , ~~~ \eta_{\lambda \mu} = \mbox{diag} (-1, 1, 1, 1) .
\end{equation}

\noindent It is convenient to denote
\begin{equation}
l_\lambda = l_0 n_\lambda , ~~~ n_\lambda = ( -1 , \bn_\lambda ) = ( -1 , n_1 , n_2 , n_3 ) , ~~ \mbox{so that} ~~ l_\lambda l_\mu = l_0^2 n_\lambda n_\mu ,
\end{equation}

\noindent and (\ref{ds2Kerr}) corresponds to
\begin{equation}
l_0 l_\lambda = \frac{ r_g r^3}{ r^4 + a^2 x_3^2} \left ( -1 , \frac{r x_1 - a x_2}{r^2 + a^2} , \frac{r x_2 + a x_1}{r^2 + a^2} , \frac{x_3 }{r } \right ) .
\end{equation}

\noindent The metric functions (\ref{ds2Kerr}) have the only singularity on the ring
\begin{equation}
x_1^2 + x_2^2 = a^2 , ~~~ x_3 = 0 .
\end{equation}

\noindent Also note the following behaviour of $r$ and these metric functions as $x_3$ tends to $0$,
\begin{eqnarray}
r & = & \left \{ \begin{array}{ll} \frac{ \textstyle a | x_3 | }{ \textstyle \sqrt{ a^2 - x_1^2 - x_2^2 - x_3^2 } } \left [ 1 + O \left ( \frac{ \textstyle x_3^2 }{ \textstyle \xi^2 } \right ) \right ] & \mbox{at } x_1^2 + x_2^2 < a^2 \\ \sqrt{a | x_3 |} \left [ 1 + O \left ( \frac{ \textstyle | x_3 | }{ \textstyle a } \right ) \right ] & \mbox{at } x_1^2 + x_2^2 = a^2 \\  \sqrt{ x_1^2 + x_2^2 + x_3^2 - a^2 } \left [ 1 + O \left ( \frac{ \textstyle x_3^2 }{ \textstyle \xi^2 } \right ) \right ] & \mbox{at } x_1^2 + x_2^2 > a^2 \end{array} \right. , \\ \label{l0^2/r_g} l_0^2 & = & \left \{ \begin{array}{ll} \frac{ \textstyle r_g a | x_3 | }{ \textstyle ( a^2 - x_1^2 - x_2^2 - x_3^2 )^{3 / 2} } \left [ 1 + O \left ( \frac{ \textstyle x_3^2 }{ \textstyle \xi^2 } \right ) \right ] & \mbox{at } x_1^2 + x_2^2 < a^2 \\ \frac{\textstyle r_g }{ \textstyle 2 \sqrt{a | x_3 |} } \left [ 1 + O \left ( \frac{ \textstyle | x_3 | }{ \textstyle a } \right ) \right ] & \mbox{at } x_1^2 + x_2^2 = a^2 \\ \frac{ \textstyle r_g }{ \textstyle \sqrt{ x_1^2 + x_2^2 + x_3^2 - a^2 } } \left [ 1 + O \left ( \frac{ \textstyle x_3^2 }{ \textstyle \xi^2 } \right ) \right ] & \mbox{at } x_1^2 + x_2^2 > a^2 \end{array} \right. , \\ n_3 & = & \left \{ \begin{array}{ll} \sqrt{ 1 - \frac{ \textstyle x_1^2 + x_2^2 + x_3^2}{ \textstyle a^2 } } \sgn x_3 \left [ 1 + O \left ( \frac{ \textstyle x_3^2 }{ \textstyle \xi^2 } \right ) \right ] & \mbox{at } x_1^2 + x_2^2 < a^2 \\ \sqrt{ \frac{ \textstyle | x_3 | }{ \textstyle a }} \sgn x_3 \left [ 1 + O \left ( \frac{ \textstyle | x_3 | }{ \textstyle a } \right ) \right ] & \mbox{at } x_1^2 + x_2^2 = a^2 \\ \frac{ \textstyle x_3 }{ \textstyle \sqrt{ x_1^2 + x_2^2 + x_3^2 - a^2 } } \left [ 1 + O \left ( \frac{ \textstyle x_3^2 }{ \textstyle \xi^2 } \right ) \right ] & \mbox{at } x_1^2 + x_2^2 > a^2 \end{array} \right. ,
\end{eqnarray}

\noindent where $a^2 \xi^2 \equiv ( x_1^2 + x_2^2 + x_3^2 - a^2 )^2$. We observe the presence of $ | x_3 | $ and $ \sgn x_3 $ here, as well as in $n_1 $, $n_2 $, which include $ | x_3 | $ linearly through $r$. This leads to the appearance of $\delta$-function source terms in the Einstein equations. The support for the source distribution contains the singularity ring, and it is of interest to analyze the system within the discrete approach, paying attention to the neighborhood of the singularity (which was such in the continuum), including the source terms.

In the most general case, we can consider an (overdetermined) decomposition of any metric $g_{\lambda \mu}$ into some $g_{(0) \lambda \mu}$ and some $l_\lambda l_\mu$,
\begin{equation}
g_{\lambda \mu} = g_{(0) \lambda \mu} + l_\lambda l_\mu , ~~~ l^\lambda l_\lambda = 0 , ~~~ l^\lambda = g_{(0) }^{ \lambda \mu} l_\mu .
\end{equation}

\noindent The Riemann tensors $R_{ \mu \nu \rho}^\lambda$ and $R_{(0) \mu \nu \rho}^\lambda$ for the metrics $g_{\lambda \mu}$ and $g_{(0) \lambda \mu}$, respectively, are related as \cite{Chandra}
\begin{equation}
R_{ \mu \nu \rho}^\lambda = R_{(0) \mu \nu \rho}^\lambda + \nabla_\nu C^\lambda_{\mu \rho } - \nabla_\rho C^\lambda_{\mu \nu } + C^\lambda_{\sigma \nu} C^\sigma_{\mu \rho} - C^\lambda_{\sigma \rho} C^\sigma_{\mu \nu} ,
\end{equation}

\noindent $\nabla_\lambda$ is the covariant differentiation operator with respect to $g_{(0) \lambda \mu}$,
\begin{equation}
C^\lambda_{\mu \nu } = \frac{1}{2} \left [ \nabla_\mu (l_\nu l^\lambda ) + \nabla_\nu (l_\mu l^\lambda ) - \nabla^\lambda ( l_\mu l_\nu ) + l^\lambda l^\sigma \nabla_\sigma (l_\mu l_\nu ) \right ] ~~~ ( C^\lambda_{\mu \lambda } = 0 ) .
\end{equation}

\noindent This gives the Ricci tensor components,
\begin{equation}
R_{\lambda \mu } = R_{ (0) \lambda \mu } + \nabla_\nu C^\nu_{\lambda \mu } - C^\nu_{\rho \mu} C^\rho_{\lambda \nu} ,
\end{equation}

\noindent and thus the Einstein equations.

If a finite difference form of the Einstein equations is considered, we have the Kerr solution in the leading order over metric variations (in the region where such variations are small, that is, in a certain sense far from singularities), in particular, $g_{(0) \lambda \mu} = \eta_{\lambda \mu }$.

Then the vacuum equations take the form
\begin{equation}\label{dC-CC}
\nabla_\nu C^\nu_{\lambda \mu } - C^\nu_{\rho \mu} C^\rho_{\lambda \nu} = 0 .
\end{equation}

\noindent It is an overdetermined system for $l_\lambda$. Moreover, starting from the answer, note that $r_g$ as an arbitrary parameter of the solution enters $l_\lambda$ through the coefficient $\sqrt{r_g}$. Then $C^\lambda_{\mu \nu }$ is a combination of $r_g$- and $r_g^2$-terms and (\ref{dC-CC}) is a combination of $r_g^n$-terms, $n = 1, 2, 3, 4$. Due to the arbitrariness of $r_g$, such equations should be fulfilled in each order of $r_g$. In the order $r_g$, (\ref{dC-CC}) reads
\begin{equation}
\nabla_\nu \nabla_\lambda (l_\mu l^\nu ) + \nabla_\nu \nabla_\mu (l_\lambda l^\nu ) C^\nu_{\lambda \mu } - \nabla_\nu \nabla^\nu (l_\lambda l_\mu ) = 0 .
\end{equation}

In particular, the $00$- and $0k$-components, taking into account stationarity, $\nabla_0 l_\lambda = 0$, are
\begin{equation}\label{ddll=0}
\bnabla^2 ( l_0^2 ) = 0
\end{equation}

\noindent and
\begin{equation}\label{ddll-ddll=0}
\nabla_m \nabla_k ( l_0 l^m ) - \bnabla^2 ( l_0 l_k ) = 0 ,
\end{equation}

\noindent respectively, outside the source region. The divergence  $\nabla^k ( \cdot )$ of the left-hand side of (\ref{ddll-ddll=0}) is identically zero, and there are two independent equations among (\ref{ddll-ddll=0}). In principle, taking into account the condition
\begin{equation}\label{ll=0}
l^\lambda l_\lambda \equiv -l_0^2 + l_1^2 + l_2^2 + l_3^2 = 0 ,
\end{equation}

\noindent (\ref{ddll=0}), (\ref{ddll-ddll=0}) and (\ref{ll=0}) present four equations for four values $l_\lambda$. A solution to (\ref{ddll-ddll=0}) can be given as
\begin{equation}
l_0 l_k = \nabla_k \eta + \ba \times \bnabla \chi ,
\end{equation}

\noindent where $\ba = (0, 0, a )$ and $\chi$ obeys
\begin{equation}
\bnabla^2 \chi = 0
\end{equation}

\noindent (the metric (\ref{ds2Kerr}) indeed has such a form). Then, provided $l_0^2$ satisfies (\ref{ddll=0}), an equation for $\chi$ follows by substituting this $\bl$ into (\ref{ll=0}),
\begin{equation}\label{(d-eta+a-d-chi)^2=l^4}
( \bnabla \eta + \ba \times \bnabla \chi )^2 = l_0^4 .
\end{equation}

\noindent But for our purposes (analytic continuation) we need an explicit closed expression for the solution, which we have not for the nonlinear partial differential equation (\ref{(d-eta+a-d-chi)^2=l^4}).

Therefore, we consider the simple ansatz for the metric in which $l_0^2$ is subject to (\ref{ddll=0}) and the other components are related to it in the same way as in the exact solution.

The fact that $l_0^2$ from (\ref{ds2Kerr}) obeys (\ref{ddll=0}) can be evident if we represent this function as a superposition of two Newton potentials with the centers shifted along $x_3$ by the imaginary values $+i a$ and $-i a$,
\begin{eqnarray}
2 \frac{r^3}{r^4 + a^2 x_3^2} & = & \frac{1}{r + i a \cos  \theta} + \frac{1}{r - i a \cos  \theta} , \\ & & r \pm i a \cos \theta = \sqrt{x_1^2 + x_2^2 + (x_3 \pm  i a )^2} , \\ 2 \frac{r^3}{r^4 + a^2 x_3^2} & = & \frac{1}{\sqrt{x_1^2 + x_2^2 + (x_3 +  i a )^2}} + \frac{1}{\sqrt{x_1^2 + x_2^2 + (x_3 -  i a )^2}} \nonumber \\ \label{((x+iy)^2)^(-1/2)+(...-...)} & \equiv & \frac{1}{\sqrt{ ( \bx + i \by )^2 }} + \frac{1}{\sqrt{ ( \bx - i \by )^2 }} , ~~ \by = (0, 0, y_3 ) , ~~ y_3 = a \sgn y_3 ,
\end{eqnarray}

\noindent ($\bx \equiv (x_1 , x_2 , x_3 )$). This is a truncated form of the Newman-Janis trick \cite{Newman} (or, in the Kerr-Schild coordinates, \cite{Rajan}), with which one obtains the Kerr geometry from the Schwarzschild geometry. This prompts us to formulate the discretization problem as that of the analytic continuation of the discrete Schwarzschild problem (i. e., the discrete Newton potential) to complex coordinates, namely, as the problem of shifting by $\pm i a$ along some axis, which thus becomes the rotational axis.

Important is the behaviour of the metric near the singularity. In the considered coordinates, the only singularity in the continuum theory is that in the overall scale of $l_\lambda$, i. e., of $l_0$ if we take $l_0$, $\bn$ as independent metric functions. This singularity is achieved at $r \to 0$, $\cos \theta = x_3 / r \to 0$, i. e., at $(\bx + i \by)^2 \to 0$ (or at $(\bx - i \by)^2 \to 0$).

In the discrete case, the complex Newton potential $1 / \sqrt{(\bx + i \by )^2}$ in (\ref{((x+iy)^2)^(-1/2)+(...-...)}) is replaced by an unknown function $\phi (\bx )$ (close to $1 / \sqrt{(\bx + i \by )^2}$ at large distances from the singularity ring). This $\phi (\bx )$ is not singular at $r \to 0$, $\cos \theta \to 0$ (it is cut off at a sufficiently small length). This can be expressed by saying that $\phi (\bx )$ is Newton's (complex) potential, but corresponding to some effective $r_{\rm eff}$ and $\cos \theta_{\rm eff}$ that do not vanish at $r = 0$, $\cos \theta = 0$,
\begin{equation}\label{phi-r-eff-theta-eff}
\phi (\bx ) = \frac{1}{ r_{\rm eff} + i y_3 \cos \theta_{\rm eff} }, ~~ r_{\rm eff} = \Re \frac{1}{ \phi }, ~~ \cos \theta_{\rm eff} = \frac{1}{y_3 } \Im \frac{1}{ \phi } .
\end{equation}

\noindent In the continuum theory, or, approximately, at large distances from the singularity, $r_{\rm eff} = r$, $\theta_{\rm eff} = \theta$. In our simple ansatz for the metric, $l_0^2$ obeys (\ref{ddll=0}), and $\bn$ is expressed in terms of the complex potential $ \phi $ and the polar angle $ \varphi $, as in the continuum solution. This is equivalent to taking as $r$, $\theta$ their effective values $r_{\rm eff}$, $\theta_{\rm eff}$ (\ref{phi-r-eff-theta-eff}); $\varphi$ is defined by $x_2 / x_1 = \tg \varphi$. Thus, we have
\begin{eqnarray}\label{l0l0}
l_0^2 & = & r_g \Re \phi \\ \label{n1} n_1 = \frac{r x_1 - a x_2}{r^2 + a^2} & \Rightarrow & \frac{x_1 \Re \phi^{-1} - a x_2}{a \sqrt{x_1^2 + x_2^2}} \sqrt{\frac{a^2 - \left( \Im \phi^{-1} \right )^2}{a^2 + \left( \Re \phi^{-1} \right )^2}} \\ \label{n2} n_2 = \frac{r x_2 + a x_1}{r^2 + a^2} & \Rightarrow & \frac{x_2 \Re \phi^{-1} + a x_1}{a \sqrt{x_1^2 + x_2^2}} \sqrt{\frac{a^2 - \left( \Im \phi^{-1} \right )^2}{a^2 + \left( \Re \phi^{-1} \right )^2}} \\ \label{n3} n_3 = \frac{x_3 }{r } & \Rightarrow & \frac{1}{y_3 } \Im \phi^{-1} ,
\end{eqnarray}

\noindent where $\phi $ is obtained by continuation to complex coordinates,
\begin{equation}
\phi ( \bx ) = \phi_0 ( \bx + i \by ) , ~~~ \by = (0, 0, a \sgn y_3 ) ,
\end{equation}

\noindent of the solution of the finite-difference Poisson equation with a point source,
\begin{equation}\label{Laplace}
\sum_{j = 1}^3 \bDelta_j \Delta_j \phi_0 ( \bx ) = \left \{ \begin{array}{rl} 0 & \mbox{at } \bx \neq 0  \\ C & \mbox{at } \bx = 0 , \end{array} \right.
\end{equation}

\noindent such that at large distances it approximates Newton's continuum potential,
\begin{equation}
\phi_0 (\bx ) \to \frac{1}{ \sqrt{ \bx^2 } } , ~~~ \bx \to \infty .
\end{equation}

A priori, we still do not exclude the presence of vertices at which $| \Im (\phi^{-1}) | > a$ ($| \cos \theta_{\rm eff} | > 1$, which would lead to imaginary $n_1$, $n_2$), which would require modifying our ansatz in the corresponding region. If the vertices are located near the singularity (this is just the case that interests us here), $| \phi |$ is large and $| \cos \theta_{\rm eff} | \ll 1$. At large distances, the discrete solution begins to approach the continuum one, and $\cos \theta_{\rm eff}$ approaches $\cos \theta$. The most dangerous region seems to be where $\theta$ is close to $0$, in which $| \cos \theta_{\rm eff} |$ values greater than $1$ are not excluded due to lattice artifacts. In this case, it is natural to put $\bn = (0, 0, \pm 1)$ at the corresponding vertices. Note that the details of our ansatz for $\bn$ do not affect the results obtained (in the leading order over metric variations).

\section{Analytic continuation}\label{continuation}

The discrete analogue of Newton's potential has the form
\begin{equation}
\frac{1}{r} \Rightarrow \phi_0 (\bx ) = \int^{\pi / b}_{-\pi / b} \int^{\pi / b}_{-\pi / b} \int^{\pi / b}_{-\pi / b} \frac{\d^3 \bp}{(2 \pi )^3} \frac{\pi b^2 \exp (i \bp \bx ) }{\sum_{j = 1}^3 \sin^2 (p_j b / 2 )} .
\end{equation}

\noindent Evidently, passing to complex coordinates under the integral sign cannot be made because of arising an unlimitedly growing exponent of $\bp$. Let us first integrate over, say, $\d p_3$. The corresponding integral is known to be
\begin{eqnarray}
\int^{\pi / b}_{-\pi / b} \frac{\d p_3}{2 \pi} \frac{ \exp (i p_3 x_3 ) }{\sum_{j = 1}^3 \sin^2 (p_j b / 2 )} & = & \int^{\pi }_{-\pi } \frac{\d q_3}{ \pi b} \frac{\exp(i q_3 \nu)}{\ch \kappa - \cos q_3} \nonumber \\ \label{int-dp3} & = & \frac{2 \sin (\pi \nu )}{\pi b \sh \kappa } \left [ \frac{1}{ \nu } + 2 \nu \sum_{n = 1}^\infty \frac{ (-1)^n \exp (- \kappa n ) }{\nu^2 - n^2} \right ] , \\ q_j \equiv p_j b , ~~~ \nu \equiv x_3 / b , & & \ch \kappa = 3 - \cos q_1 - \cos q_2 . \nonumber
\end{eqnarray}

\noindent Now we can pass to complex $x_3$, so $\Im \nu = \pm a / b$. But this leads to a non-physically exponentially large right-hand side of (\ref{int-dp3}) at a small length scale $b$ due to the presence of $\sin ( \pi \nu )$ there. On the other hand, it is admissible to add to (\ref{int-dp3}) a function of $\nu$, which vanishes for integer $\nu$. We add to the sum in (\ref{int-dp3}) an analytic function $ - f ( \nu )$ that has no poles for integer $\nu $. We denote the (modified) sum in (\ref{int-dp3}) multiplied by $\sin ( \pi \nu ) / \pi$ as
\begin{equation}
g ( \nu ) \equiv \left [ \frac{1}{ \nu } + 2 \nu \sum_{n = 1}^\infty \frac{ (-1)^n \exp (- \kappa n ) }{\nu^2 - n^2} \right ] \frac{\sin ( \pi \nu )}{\pi } - f (\nu ) \frac{\sin ( \pi \nu )}{\pi } .
\end{equation}

\noindent For an integer $\nu $, we have
\begin{equation}
g ( \nu ) = \exp ( - \kappa | \nu | ) , ~~~ | \nu | = 0, 1, 2, \dots .
\end{equation}

\noindent Continuation to complex $\nu $ gives
\begin{equation}
g ( \nu ) = \exp ( - \kappa \sqrt{ \nu^2 } ) , ~~~ \sqrt{ \nu^2 } = \nu \sgn \Re \nu ,
\end{equation}

\noindent where $\sqrt{ \nu^2 }$ is defined by $\sqrt{1} = 1$ on the complex plane with the cut at $\nu^2 \in ( - \infty , 0 ]$, that is, along the imaginary axis $\Re \nu = 0$. This means
\begin{equation}
\int^{\pi / b}_{-\pi / b} \frac{\d p_3}{2 \pi} \frac{ \exp (i p_3 x_3 ) }{\sum_{j = 1}^3 \sin^2 (p_j b / 2 )} \Rightarrow \frac{2}{ b \sh \kappa } \exp \left ( - \kappa \frac{x_3 + i y_3}{b } \sgn x_3 \right ) \mbox{ at } x_3 \Rightarrow x_3 + i y_3
\end{equation}

\noindent for the analytic continuation of (\ref{int-dp3}) and implies
\begin{equation}
f ( \nu ) = \frac{1 - g ( \nu )}{ \nu } + 2 \nu \sum_{n = 1}^\infty \frac{( - 1 )^n [ \exp ( - \kappa  n ) - g ( \nu ) ] }{\nu^2 - n^2} ,
\end{equation}

\noindent so $f ( \nu ) $ indeed has no poles for integer $\nu $. This uses the partial fraction expansion for sine,
\begin{equation}
\frac{\pi }{\sin ( \pi \nu )} = \frac{1}{ \nu } +2 \nu \sum_{n = 1}^\infty \frac{ ( - 1 )^n}{\nu^2 - n^2} .
\end{equation}

In fact, this corresponds to calculating the integral (\ref{int-dp3}) by taking only the residue at the pole inside the contour formed by the segments $[ - \pi , + \pi ]$, $[ - \pi , - \pi +i L ]$, $[ - \pi + i L , + \pi + i L ]$, $[ + \pi , + \pi +i L ]$, $L \to \infty \cdot \sgn \nu $, in the complex $q$ plane without taking into account the contributions from the last three segments, $[ - \pi , - \pi +i L ]$, $[ - \pi + i L , + \pi + i L ]$, $[ + \pi , + \pi +i L ]$. This does not change the integral at physical (integer) points, but removes the aforementioned non-physical (infinite, as b tends to zero) terms.

The uniqueness of such a continuation from the discrete set of points (of course, on the class of analytic functions) is based on some discussion of the fact that any nontrivial addition to $g ( \nu )$, which vanishes for integer $\nu $, must contain a factor that is the product of the factors $( 1 - \nu / n )$, $n = \pm 1, \pm 2, \dots$, and $\nu $ (Weierstrass factorization theorem), which leads to $\sin ( \pi \nu )$,
\begin{equation}
\nu \prod_{n = 1}^\infty \left ( 1 - \frac{\nu^2 }{n^2 } \right ) = \frac{\sin ( \pi \nu )}{\pi } ,
\end{equation}

\noindent exponentially diverging at $| \Im \nu | = a / b \to \infty$ at $b \to 0$. This uses the product expansion for sine.

Now we can pass to complex coordinates under the integral sign. So far, we can do this without restricting the range of admissible values of the real parts of the coordinates (${\rm I \!\! R }^3$) only for $x^3$: if we pass to complex $x_1$, $x_2$,
\begin{equation}
x_1 \Rightarrow x_1 + i y_1 , ~~~ x_2 + i y_2 ,
\end{equation}

\noindent then for $x_3^2 < y_1^2 + y_2^2$ the integrand grows exponentially in a certain region. That is, starting with the integration over $\d p_3$, we can relatively easily analyze the case of rotation around $x_3$.

Thus, we have
\begin{eqnarray}
\phi ( \bx ) & = & \phi_0 ( x_1 , x_2 , x_3 + i y_3 ) \nonumber \\ & = & \int_{- \pi }^\pi \int_{- \pi }^\pi \frac{ \d q_1 \d q_2 }{2 \pi b \sh \kappa } \exp \left ( i q_1 \frac{ x_1 }{ b } + i q_2 \frac{ x_2 }{ b } - \kappa \frac{x_3 + i y_3}{ b } \sgn x_3 \right ) ,
\end{eqnarray}

\noindent $y_3 = \pm a = a \sgn y_3$. By redefining $q_j \Rightarrow q_j \sgn (x_3 y_3 )$, we rewrite this as
\begin{equation}\label{phi-discr}
\phi ( \bx ) = \int_{- \pi }^\pi \int_{- \pi }^\pi \frac{ \d q_1 \d q_2 }{2 \pi b \sh \kappa } \exp \left \{ i \frac{ s }{ b } \left [ q_1 x_1 + q_2 x_2 - a \kappa \right ] - \frac{ | x_3 | }{ b } \kappa \right \},
\end{equation}

\noindent $s \equiv \sgn (x_3 y_3 )$.

\section{The metric in the discrete case}\label{metric}

In the continuum limit, $b \to 0$, we return from $q_j$ to $p_j = q_j / b$ and note that generally (not at singular points) the cut-off effect of the exponent $\exp ( i \bp \bx )$ leads to limiting $p_j$, not $q_j$ (say, by $1 / r$ if $r \gg a$). Then, in this limit, $\kappa = b \sqrt{p_1^2 + p_2^2}$, and after passing to polar variables, $p_1 = \rho \cos \varphi$, $p_2 = \rho \sin \varphi$, the integral (\ref{phi-discr}) gives the continuum expression for the analytically continued Newton potential,
\begin{eqnarray}\label{phi-cont}
\phi & = & \int_0^{2 \pi } \frac{\d \varphi }{2 \pi } \int_0^\infty \d \rho \exp \left \{ \rho \left [ i s \left ( x_1 \cos \varphi + x_2 \sin \varphi - a \right ) - | x_3 | \right ] \right \} \nonumber \\ & = & \int_0^{2 \pi } \frac{\d \varphi }{2 \pi } \frac{1}{| x_3 | - i s \left ( \sqrt{x_1^2 + x_2^2 } \cos \phi - a \right )} = \frac{1}{\sqrt{ \left ( | x_3 | + i s a \right )^2 + x_1^2 + x_2^2 }} \nonumber \\ & = & \frac{1}{\sqrt{ ( \bx + i \by )^2 }}.
\end{eqnarray}

\noindent Here, the cut-off effect of the exponent and, therefore, the applicability of the continuum expression is determined, roughly, by $ \max \left ( \left | \sqrt{x_1^2 + x_2^2} - a \right | , \left | x_3 \right | \right )$, that is, by the distance from the singularity ring. In relation to what should this distance be large? The answer to this question, as a by-product, will follow from the analysis in the neighborhood of the singularity ring, to which we now turn.

To this end, we pass to the variables $\lambda_j = \sin (q_j / 2)$, $j = 1, 2$, then the integral (\ref{phi-discr}) takes the form
\begin{eqnarray}
\phi & = & \int_{- 1}^{+ 1} \int_{- 1}^{+ 1} \frac{ \d \lambda_1 \d \lambda_2 }{ \pi b } \left [ \left ( 1 - \lambda_1^2 \right ) \left (1 - \lambda_2^2 \right ) \left ( 1 + \lambda_1^2 + \lambda_2^2 \right ) \left ( \lambda_1^2 + \lambda_2^2 \right ) \right ]^{- 1 / 2} \vphantom{\left [ 2 \sqrt{\left ( 1 + \lambda_1^2 + \lambda_2^2 \right ) \left ( \lambda_1^2 + \lambda_2^2 \right )} \right ]} \nonumber \\ & & \cdot \exp \left \{ \frac{ 1 }{ b } \left [ 2 i s \left ( x_1 \arcsin \lambda_1 + x_2 \arcsin \lambda_2 \right ) \vphantom{\left [ 2 \sqrt{\left ( 1 + \lambda_1^2 + \lambda_2^2 \right ) \left ( \lambda_1^2 + \lambda_2^2 \right )} \right ]} \right. \right. \nonumber \\ & & \left. \left. - \left ( i s a + | x_3 | \right ) \arsh \left [ 2 \sqrt{\left ( 1 + \lambda_1^2 + \lambda_2^2 \right ) \left ( \lambda_1^2 + \lambda_2^2 \right )} \right ] \right ] \right \} .
\end{eqnarray}

\noindent We pass to polar coordinates for the integration variables $\lambda_1$, $\lambda_2$ and for the point $x_1$, $x_2$,
\begin{equation}
\left. \begin{array}{rcl} \lambda_1 & = & \Lambda \cos \varphi  \\ \lambda_2 & = & \Lambda \sin \varphi \end{array} \right \} , ~~~ \left. \begin{array}{rcccl} -1 & \leq & \Lambda \cos \varphi & \leq & 1  \\ -1 & \leq & \Lambda \sin \varphi & \leq & 1 \end{array} \right \} {\cal D} , ~~~ \left. \begin{array}{rcl} x_1 & = & r_0 \cos \varphi_0  \\ x_2 & = & r_0 \sin \varphi_0 \end{array} \right \} ,
\end{equation}

\noindent so that
\begin{eqnarray}
& & \hspace{-5mm} \phi = \iint_{\cal D} \frac{\d \Lambda \d \varphi }{ \pi b } \left [ 1 - \Lambda^4 + \left ( \Lambda^4 + \Lambda^6 \right ) \cos^2 \varphi \sin^2 \varphi \right ]^{- 1 / 2} \exp \left \{ \frac{ 1 }{ b } \left [ 2 i s r_0 \vphantom{\left ( 2 \Lambda \sqrt{1 + \Lambda^2} \right )} \left [ \cos \varphi_0 \vphantom{\left ( \Lambda \cos \varphi \right )} \right. \right. \right. \nonumber \\ & & \hspace{-10mm} \left. \left. \left. \cdot \arcsin \left ( \Lambda \cos \varphi \right ) + \sin \varphi_0 \arcsin \left ( \Lambda \sin \varphi \right ) \right ] - \left ( i s a + \left | x_3 \right | \right ) \arsh \left ( 2 \Lambda \sqrt{1 + \Lambda^2} \right ) \right ] \vphantom{\frac{ 1 }{ b }} \right \} .
\end{eqnarray}

\noindent For small $b$, a contribution to this integral dominates from the region in which the functions arcsin and arsh in the exponent are small, that is, the values of $\Lambda$ are small enough to at least lie in the convergence domain of the Taylor series for the exponent in the integration region ${\cal D}$ (this refers to the function arsh, for which the Taylor series converges at $2 \Lambda \sqrt{1 + \Lambda^2} \leq 1$, that is, at $\Lambda \leq 1 / \sqrt{2 \left ( \sqrt{2} + 1 \right )}$). Making such a Taylor series expansion, we obtain
\begin{eqnarray}\label{phi-small-b}
& & \hspace{-4mm} \phi = \iint_{\cal D} \frac{\d \Lambda \d \varphi }{ \pi b } \left [ 1 - \Lambda^4 + \left ( \Lambda^4 + \Lambda^6 \right ) \cos^2 \varphi \sin^2 \varphi \right ]^{- 1 / 2} \exp \left \{ \frac{ 1 }{ b } \left [ 2 i s \vphantom{\left ( \Lambda^5 \right )} \right. \right. \nonumber \\ & & \hspace{-4mm} \cdot \Lambda \left [ r_0 \cos (\varphi - \varphi_0 ) - a \right ] + 3^{- 1} i s \Lambda^3 \left [ r_0 \left ( \cos \varphi_0 \cos^3 \varphi + \sin \varphi_0 \sin^3 \varphi \right ) + a \right ] - \left | x_3 \right | \left ( 2 \Lambda \right. \nonumber \\ & & \hspace{-6mm} \left. \left. \left. - 3^{- 1} \Lambda^3 \right ) + i s r_0 \cos \varphi_0 O \left ( \Lambda^5 \cos^5 \varphi \right ) + i s r_0 \sin \varphi_0 O \left ( \Lambda^5 \sin^5 \varphi \right ) + \left | x_3 \right | O \left ( \Lambda^5 \right ) \right ] \vphantom{\frac{ 1 }{ b }} \right \} .
\end{eqnarray}

\noindent It is seen that $\phi$ can sharply increase at $r_0 = a$, $x_3 = 0$ due to a qualitative decrease in the suppression effect by the exponential at these points. Thus, this is a refined position of the (regularized) singularity. At these points, the linear term in $\Lambda$ in the expansion of the exponent over $\Lambda$, $\varphi - \varphi_0$ vanishes, and the contribution of large $\Lambda$, $\varphi - \varphi_0$ is suppressed by the next cubic terms $\Lambda^3$ and $\Lambda ( \varphi - \varphi_0 )^2 $. As a result, the dominant contribution comes from the typical values of $\Lambda$, $\varphi - \varphi_0$ defined by $a \Lambda^3 \sim b$, $a \Lambda (\varphi - \varphi_0 )^2 \sim b$, that is, $\Lambda \sim ( b/ a )^{1 / 3}$, $|\varphi - \varphi_0 | \sim ( b/ a )^{1 / 3}$.

Thus, the typical values of $\Lambda$, $\varphi - \varphi_0$ are proportional to $\sqrt[3]{b}$. Therefore, $\sqrt[3]{b}$ is a natural small parameter. Then we consider a neighborhood of the established analogue of the singularity ring and write the coordinates in the form
\begin{equation}
\left. \begin{array}{rcr} x_1 & = & a \cos \varphi_0 + k_1 b  \\ x_2 & = & a \sin \varphi_0 + k_2 b \\ x_3 & = & k_3 b \end{array} \right \} ,
\end{equation}

\noindent where $k_1$, $k_2$, $k_3$ are integers of order $1$, and expand in (fractional) powers of $b$, attributing $b^{1 / 3}$ to $\Lambda$ and $\varphi - \varphi_0$. The expression for $\phi$ reads
\begin{eqnarray}
\phi & = & \iint_{\cal D} \frac{\d \Lambda \d \varphi }{\pi b } \exp \left ( \frac{i s }{ b } F \right ) , ~~ F = -a \Lambda ( \varphi - \varphi_0 )^2 + \frac{1}{3} a \left ( 1 + \cos^4 \varphi_0 + \sin^4 \varphi_0 \right ) \Lambda^3 \nonumber \\ & & - \frac{1}{4} a \sin ( 4 \varphi_0 ) \Lambda^3 \left ( \varphi - \varphi_0 \right ) + O \left [ \Lambda ( \varphi - \varphi_0 )^4 , \Lambda^3 ( \varphi - \varphi_0 )^2 , \Lambda^5 \right ] \nonumber \\ & & + 2 b \Lambda \left ( k_1 \cos \varphi_0 + k_2 \sin \varphi_0 + i s | k_3 | \right ) + O \left [ b \Lambda ( \varphi - \varphi_0 ) , b \Lambda^4 \right ] .
\end{eqnarray}

\noindent Shown are $F$ terms of order $b$ ($\Lambda ( \varphi - \varphi_0 )^2$ and $\Lambda^3$) and $b^{4 / 3}$ ($\Lambda^3 ( \varphi - \varphi_0 )$ and $b \Lambda$). The subsequent terms $O ( \dots )$ are $\sim b^{5 / 3}$ and less. In particular, the dependence of the denominator on $\Lambda$ in (\ref{phi-small-b}), when rewritten in the exponential form, adds $O \left ( b \Lambda^4 \right ) = O \left ( b^{7 / 3} \right )$ to $F$.

After integration over $\varphi$, we get
\begin{eqnarray}
\phi = \exp \left ( - \frac{i s \pi}{4} \right ) \frac{2}{\sqrt{\pi a b}} \sqrt[6]{ \frac{3 b}{a \mu} } \int\limits_0^\infty \d w \exp \left [ i w^6 + 2 \sqrt[3]{ \frac{3 b}{a \mu} } \left ( i s k_{12} - | k_3 | \right ) w^2 \right ] , \phantom{\phi} \\ \mu = 1 + \cos^4 \varphi_0 + \sin^4 \varphi_0 , k_{1 2} = k_1 \cos \varphi_0 + k_2 \sin \varphi_0 , \Lambda = (3 b / (a \mu))^{1 / 3} w^2 . \phantom{\phi \phi} \nonumber
\end{eqnarray}

\noindent With the help of contour integration in the complex plane of $w$, we pass from the oscillating exponent to a monotonic one (mainly) by a complex rotation of the variable, $w = u \exp ( i s \pi / 12)$,
\begin{equation}
\phi = \exp \left ( - \frac{i s \pi}{6} \right ) \frac{2}{\sqrt{\pi a b}} \sqrt[6]{ \frac{3 b}{a \mu} } \int\limits_0^\infty \d u \exp \left [ - u^6 + 2 \sqrt[3]{ \frac{3 b}{a \mu} } \exp \left ( \frac{i s \pi}{6} \right ) \left ( i s k_{12} - | k_3 | \right ) u^2 \right ] .
\end{equation}

\noindent First consider the case
\begin{equation}\label{b^1/3|k|>>1}
(b / a)^{1 / 3} \left | i k_{12} - | k_3 | \right | \gg 1 .
\end{equation}

\noindent Then we can neglect the $u^6$ term in the exponent and find
\begin{equation}\label{phi-cont-near-ring}
\phi = \frac{1}{ \sqrt{ 2a \left [ x_{1 2} + i s | x_3 | \right ] } } , ~~~ x_{1 2} \equiv b k_{1 2} = x_1 \cos \varphi_0 + x_2 \sin \varphi_0 .
\end{equation}

\noindent If we optimally choose a point on the ring $(a \cos \varphi_0 , a \sin \varphi_0 , 0)$, relative to which the coordinates $(b k_1 , b k_2 , b k_3 )$ of the current vertex are determined, so that $\varphi_0$ is the polar angle of $\bx$, then
\begin{equation}
x_{1 2} = ( x_1 - a \cos \varphi_0 ) \cos \varphi_0 + ( x_2 - a \sin \varphi_0 ) \sin \varphi_0 = \sqrt{x_1^2 + x_2^2} -a .
\end{equation}

\noindent At
\begin{equation}\label{max<<a}
\max \left ( \left | \sqrt{x_1^2 + x_2^2} - a \right | , | x_3 | \right ) \ll a ,
\end{equation}

\noindent we have
\begin{equation}
2a \left [ x_{1 2} + i s | x_3 | \right ] \approx x_1^2 +x_2^2 + \left ( i s a + | x_3 | \right )^2 = ( \bx + i \by )^2
\end{equation}

\noindent and thus (\ref{phi-cont-near-ring}) reproduces the continuum function (\ref{phi-cont}) near the ring. In principle, the exact continuum answer for all points would follow if the above integration over $\varphi$ would be exact (giving some Bessel function): at sufficiently large distances from the ring, the dominating contribution to the integral is provided by smaller values of $\Lambda$, the latter enters the coefficient at $1 - \cos ( \varphi - \varphi_0 )$, therefore, considerable values of the angle $\varphi - \varphi_0$ can give a significant contribution, so the quadratic approximation of $1 - \cos ( \varphi - \varphi_0 )$ used above is not suitable if (\ref{max<<a}) is not satisfied.

Thus, the answer becomes the continuum one at the distances satisfying (\ref{b^1/3|k|>>1}) or
\begin{equation}\label{dist>>a^1/3b^2/3}
\max \left ( \left | \sqrt{x_1^2 + x_2^2} - a \right | , | x_3 | \right ) \gg \sqrt[3]{a b^2} .
\end{equation}

\noindent This estimate is stronger than the simplest one available, such as in the Schwarzschild case, which states that the distances should be much larger than $b$ itself.

In the vicinity of the ring, we expand over $i s k_{1 2} - | k_3 |$,
\begin{eqnarray}
\phi & = & \frac{1}{ a \sqrt{3 \pi \mu } } \left [ \Gamma \left ( \frac{1}{6} \right ) \sqrt[3]{ \frac{a \mu}{3 b} } \exp \left ( - \frac{i s \pi }{6} \right ) + 2 \sqrt{\pi } \left ( i s k_{1 2} - | k_3 | \right ) \right. \nonumber \\ & & \left. + \frac{4 \pi}{ \Gamma ( 1 / 6 )} \sqrt[3]{ \frac{3 b}{a \mu}} \exp \left ( \frac{i s \pi}{6} \right ) \left ( i s k_{1 2} - | k_3 | \right )^2 \right. \nonumber \\ & & \left. + \frac{2 }{9 } \Gamma (1 / 6 ) \left ( \frac{3 b}{a \mu} \right )^{2 / 3} \exp \left ( \frac{i s \pi}{3} \right ) \left ( i s k_{1 2} - | k_3 | \right )^3 + \dots \right ] , \\
\frac{1 }{ \phi } & = & a \sqrt{3 \pi \mu } \left \{ \frac{1}{ \Gamma ( 1 / 6 )} \sqrt[3]{ \frac{3 b}{a \mu}} \exp \left ( \frac{i s \pi}{6} \right ) \right. \nonumber \\ & & \left. - \frac{2 \sqrt{ \pi }}{\left ( \Gamma ( 1 / 6 ) \right )^2} \left ( \frac{3 b}{a \mu} \right )^{2 / 3} \exp \left ( \frac{i s \pi}{3} \right ) \left ( i s k_{1 2} - | k_3 | \right ) + 0 \cdot \left ( i s k_{1 2} - | k_3 | \right )^2 \right. \nonumber \\ & & \left. + \left [ \frac{8 \pi^{3 / 2}}{\left ( \Gamma ( 1 / 6 ) \right )^4} - \frac{2}{9 \Gamma (1 / 6)} \right ] \left ( \frac{3 b}{a \mu} \right )^{4 / 3} \exp \left ( \frac{2 i s \pi}{3} \right ) \left ( i s k_{1 2} - | k_3 | \right )^3 + \dots \right \} , \hspace{5mm}
\end{eqnarray}

\noindent so the real and imaginary parts in the orders of interest are
\begin{eqnarray}\label{Re-phi}
\Re \phi & = & \frac{1}{ 2 a \sqrt{3 \pi \mu } } \left [ \sqrt{3 } \Gamma \left ( \frac{1}{6} \right ) \sqrt[3]{ \frac{a \mu}{3 b} } - 4 \sqrt{ \pi } | k_3 | \right. \nonumber \\ & & \left. + \frac{4 \pi}{ \Gamma ( 1 / 6 )} \sqrt[3]{ \frac{3 b}{a \mu}} \left ( - \sqrt{3} k_{1 2}^2 + \sqrt{3} k_3^2 + 2 k_{1 2} | k_3 | \right ) + \dots \right ] , \\
\frac{1 }{ a } \Re \frac{1 }{ \phi } & = & \frac{ \sqrt{3 \pi \mu } }{2 \Gamma (1 / 6)} \left [ \sqrt[3]{\frac{3 b}{ a \mu }} \sqrt{3 } + \frac{2 \sqrt{ \pi }}{ \Gamma (1 / 6) } \left ( \frac{3 b}{ a \mu } \right )^{2 / 3} \left ( k_{1 2} \sqrt{3 } - | k_3 | \right ) \right. \nonumber \\ & & \left. + 0 \cdot k_{1 2}^2 + 0 \cdot k_{1 2}| k_3 | + 0 \cdot k_3^2 + \dots \vphantom{\left ( \frac{3 b}{ a \mu } \right )^{2 / 3}} \right ] , \\
\label{Im-1/phi} \frac{1 }{ y_3 } \Im \frac{1 }{ \phi } & = & \frac{ \sqrt{3 \pi \mu } }{2 \Gamma (1 / 6)} \left [ \sqrt[3]{\frac{3 b}{ a \mu }} \sgn k_3 - \frac{2 \sqrt{ \pi }}{ \Gamma (1 / 6) } \left ( \frac{3 b}{ a \mu } \right )^{2 / 3} \left ( k_{1 2} \sgn k_3 + k_3 \sqrt{3 } \right ) \right. \nonumber \\ & & \left. + 0 \cdot k_{1 2}^2 + 0 \cdot k_{1 2}| k_3 | + 0 \cdot k_3^2 + \dots \vphantom{\left ( \frac{3 b}{ a \mu } \right )^{2 / 3}} \right ] .
\end{eqnarray}

\section{Riemann tensor}\label{riemann}

The metric functions obtained have kinks and breaks at $x_3 = 0$ (the extension of the functions from integer to arbitrary coordinates is implied). Consider first the case $x_3 \neq 0 $. We substitute the metric (\ref{l0l0}) - (\ref{n3}) with $\phi$ found, (\ref{Re-phi}) - (\ref{Im-1/phi}), into (the discrete form of) the fully covariant Riemann tensor,
\begin{eqnarray}
R_{\lambda \mu \nu \rho} & = & \frac{1}{2 b^2} \left [ \Delta_\mu \Delta_\nu \left ( l_\lambda l_\rho \right ) + \Delta_\lambda \Delta_\rho \left ( l_\mu l_\nu \right ) - \Delta_\mu \Delta_\rho \left ( l_\lambda l_\nu \right ) - \Delta_\lambda \Delta_\nu \left ( l_\mu l_\rho \right ) \right ] \nonumber \\ & & + \left ( \eta^{\sigma \tau} - l^\sigma l^\tau \right ) \left ( C_{\sigma , \mu \nu } C_{\tau , \lambda \rho } - C_{\sigma , \mu \rho } C_{\tau , \lambda \nu } \right ) , \\
C_{\lambda , \mu \nu } & = & \frac{1}{2 b} \left [ \Delta_\mu \left ( l_\nu l_\lambda \right ) + \Delta_\nu \left ( l_\lambda l_\mu \right ) - \Delta_\lambda \left ( l_\mu l_\nu \right ) \right ] .
\end{eqnarray}

\noindent Acting by the finite-difference operators $\Delta_\lambda$, $\Delta_\lambda \Delta_\mu$ on the metric in the neighborhood of the singularity ring, we consider the minimal order in the expansion of the metric over $\bk$, which does not lead to zero, linear for $\Delta_\lambda$ and bilinear for $\Delta_\lambda \Delta_\mu$, since the expansion proceeds effectively over $\sqrt[3]{b} \bk$, and higher orders are suppressed by powers of $\sqrt[3]{b}$. And on polynomials of such minimal orders, the finite differences act as derivatives. For the found expansion of the metric for orders in $b$ for finite differences, we have
\begin{eqnarray}\label{Delta-on-functions-orders}
& & \Delta_\lambda \Re \phi = O (1) \mbox{ (at $\lambda = 3$), } ~ \Delta_\lambda \Delta_\mu \Re \phi \sim b^{1 / 3} , ~~ \Delta_\lambda n_\mu \sim b^{2 / 3} , \nonumber \\ & & \Delta_\lambda \Delta_\mu n_\nu \sim b^{5 / 3} .
\end{eqnarray}

\noindent The last formula takes into account the dependence of $n_1 , n_2$ (\ref{n1}), (\ref{n2}) both on $b^{2 / 3} k_1$, $b^{2 / 3} k_2$ due to our expansion of $\phi$, and on $b k_1$, $b k_2$ through the already present dependence on $x_1$, $x_2$ in addition to our expansion. Note also that the metric effectively depends on $k_1$, $k_2$ via $\sqrt[3]{ b } k_{1 2} = \sqrt[3]{ b } \left ( k_1 \cos \varphi_0 + k_2 \sin \varphi_0 \right )$ in our expansion; the already existing dependence of $n_1 , n_2$ on $x_1$, $x_2$ leads to a weaker (by a power of $b$) dependence on $k_1$, $k_2$ through $b k_1$, $b k_2$. Therefore,
\begin{equation}
n^{( 0 ) \lambda} \Delta_\lambda \left ( l_\mu l_\nu \right ) \sim b , \mbox{ where } n^{( 0 ) \lambda} = ( 1 , - \sin \varphi_0 , \cos \varphi_0 , 0 )
\end{equation}

\noindent is the value of $n^\lambda$ on the singularity ring ($\bk = 0$) in the absence of discreteness corrections ($b \to 0$). $n^\lambda$ itself on the ring differs from $n^{( 0 ) \lambda}$ by $O \left ( b^{1 / 3} \right )$, so
\begin{equation}\label{nDelta-on-functions-orders}
n^\lambda \Delta_\lambda \Re \phi \sim b^{1 / 3} , ~~~ n^\lambda \Delta_\lambda n_\mu \sim b , ~~~ n^\lambda \Delta_\lambda \left ( l_\mu l_\nu \right ) \sim b^{1 / 3} .
\end{equation}

\noindent With such estimates, the leading contribution to $R_{\lambda \mu \nu \rho}$ can be distinguished at $b \to 0$. The general form of the term with the second finite difference is as follows:
\begin{eqnarray}
& & \Delta_\mu \Delta_\nu \left [ ( \Re \phi ) n_\lambda n_\rho \right ] = \left ( \Delta_\mu \Delta_\nu \Re \phi \right ) n_\lambda n_\rho \nonumber \\& & + \left [ \left ( \Delta_\mu \Re \phi \right ) \Delta_\nu \left ( n_\lambda n_\rho \right ) + \left ( \Delta_\nu \Re \phi \right ) \Delta_\mu \left ( n_\lambda n_\rho \right ) \right ] \nonumber \\& & + ( \Re \phi ) \Delta_\mu \Delta_\nu \left ( n_\lambda n_\rho \right ) = O \left ( b^{1 / 3} \right ) + O \left ( b^{2 / 3} \right ) + O \left ( b \right ) ,
\end{eqnarray}

\noindent with orders for the resulting three terms, respectively, of which the first is the largest,
\begin{equation}
\left ( \Delta_\mu \Delta_\nu \Re \phi \right ) n_\lambda n_\rho \sim b^{1 / 3} .
\end{equation}

\noindent All other terms in $R_{\lambda \mu \nu \rho}$, including those that are bilinear in $C_{\lambda , \mu \nu }$, are smaller (have higher orders in $b$). Roughly speaking, finite differences should be concentrated on $\Re \phi$. As a result, the main contribution to $R_{\lambda \mu \nu \rho}$ at $x_3 \neq 0$, $b \to 0$ takes the form
\begin{eqnarray}\label{Rlmnr-on-ring}
R_{\lambda \mu \nu \rho} & = & \frac{r_g}{ 2 b^2 } \left ( n_\lambda n_\rho \Delta_\mu \Delta_\nu \Re \phi + n_\mu n_\nu \Delta_\lambda \Delta_\rho \Re \phi \right. \nonumber \\ & & \left. - n_\mu n_\rho \Delta_\lambda \Delta_\nu \Re \phi - n_\lambda n_\nu \Delta_\mu \Delta_\rho \Re \phi \right ) \sim b^{- 5 / 3} .
\end{eqnarray}

In the neighborhood of $x_3 = 0$, it is appropriate to use the more accurate Hermitian definition
\begin{equation}
- \frac{1}{2} \left ( \bDelta_\lambda \Delta_\mu + \bDelta_\mu \Delta_\lambda \right ) \equiv - \bDelta_{ ( \lambda } \Delta_{ \mu ) }
\end{equation}

\noindent instead of $\Delta_\lambda \Delta_\mu$ for the discrete version of $\partial_\lambda \partial_\mu$, similar to the definition used in the discrete Laplacian (\ref{Laplace}). Otherwise, calculating $\Delta_3 \Delta_3 f( k_3 )$ for some metric function $f( k_3 )$,
\begin{equation}
\Delta_3 \Delta_3 f( k_3 ) = f( k_3 ) - 2 f( k_3 - 1 ) + f( k_3 - 2) ,
\end{equation}

\noindent at $k_3 = 1$, will require knowledge of $f( - 1 )$, that is, passing through the non-analyticity point $k_3 = 0$.

In the context of our calculation, we need to contract two vectors of the type $\Delta_\lambda A$, $\Delta_\mu B$, say, when finding $C_{\sigma , \mu \nu} C_{\lambda \rho }^\sigma$ above or $R_{\lambda \mu \nu \rho} R^{\lambda \mu \nu \rho}$ below (if instead of $\Delta_\lambda A$ and/or $\Delta_\mu B$ we have a value proportional to $l_\lambda$ and/or $l_\mu$, the contraction procedure is trivial and is done using the $\eta^{\lambda \mu}$ metric). We have
\begin{eqnarray}
& & g^{\lambda \mu } ( \Delta_\lambda A ) \Delta_\mu B = \eta^{\lambda \mu } ( \Delta_\lambda A ) \Delta_\mu B - l^\lambda ( \Delta_\lambda A ) l^\mu \Delta_\mu B = \eta^{\lambda \mu } ( \Delta_\lambda A ) \Delta_\mu B \nonumber \\ & & - ( \Re \phi ) ( n^\lambda \Delta_\lambda A ) n^\mu \Delta_\mu B = \eta^{\lambda \mu } ( \Delta_\lambda A ) ( \Delta_\mu B ) \left ( 1 + O \left ( b^{1 / 3} \right ) \right )
\end{eqnarray}

\noindent according to the above estimates for $n^\lambda \Delta_\lambda$, whose action introduces an additional smallness $b^{1 / 3}$ in comparison with one $\Delta_\lambda$ (acting on the functions of $k_3$ and mainly $k_{1 2}$) and for $\Re \phi = O \left ( b^{- 1 / 3 } \right )$. Thus, in the leading order, the contraction is performed by $\eta^{\lambda \mu }$.

The found leading contribution to $R_{\lambda \mu \nu \rho}$ in the neighborhood of the singularity ring at $x_3 \neq 0$ (\ref{Rlmnr-on-ring}) can be written in the form of three matrices $A$, $B$, $C$:
\begin{equation}
R_{0 k 0 l} = A_{k l} , ~~~ R_{k l m n} = \epsilon_{k l p} \epsilon_{m n r} C_{p r} , ~~~R_{0 k l m} = \epsilon_{l m n} B_{n k} ,
\end{equation}

\noindent of which $A$, $C$ are symmetric. These matrices are
\begin{eqnarray}
A = - C = \left ( \begin{array}{lll} \sqrt{3} \cos^2 \varphi_0 & \sqrt{3} \cos \varphi_0 \sin \varphi_0 & -\sgn x_3 \cos \varphi_0 \\ \sqrt{3} \cos \varphi_0 \sin \varphi_0 & \sqrt{3} \sin^2 \varphi_0 & -\sgn x_3 \sin \varphi_0 \\ -\sgn x_3 \cos \varphi_0 & -\sgn x_3 \sin \varphi_0 & - \sqrt{3} \end{array} \right ) , \\
B = \left ( \begin{array}{lll} \sgn x_3 \cos^2 \varphi_0 & \sgn x_3 \cos \varphi_0 \sin \varphi_0 & \sqrt{3} \cos \varphi_0 \\ \sgn x_3 \cos \varphi_0 \sin \varphi_0 & \sgn x_3 \sin^2 \varphi_0 & \sqrt{3} \sin \varphi_0 \\ \sqrt{3} \cos \varphi_0 & \sqrt{3} \sin \varphi_0 & - \sgn x_3 \end{array} \right ) .
\end{eqnarray}

\noindent They have the properties
\begin{equation}\label{Rlm=0}
A_{k l} = - C_{k l} , ~~~ A_{k k} = 0 , ~~~ B_{k l} = B_{l k} ,
\end{equation}

\noindent which are equivalent to the vacuum Einstein equations $R_{\lambda \mu} = 0$, and $B_{k k} = 0$, which is equivalent to the identity $R_{0 1 2 3} + R_{0 2 3 1} + R_{0 3 1 2} = 0$. $A$ and $B$ can be combined into one complex $D$,
\begin{equation}
D = A + i B = \left ( \sqrt{ 3 } + i \sgn x_3 \right ) \left ( \begin{array}{lll} \cos^2 \varphi_0 & \cos \varphi_0 \sin \varphi_0 & i \cos \varphi_0 \\ \cos \varphi_0 \sin \varphi_0 & \sin^2 \varphi_0 & i \sin \varphi_0 \\ i \cos \varphi_0 & i \sin \varphi_0 & - 1 \end{array} \right ) .
\end{equation}

\noindent Two complex eigenvalues of $D$ (the third eigenvalue is equal to minus the sum of these two due to vanishing of the trace of $D$) provide information about four independent curvature invariants. So, all eigenvalues are equal to zero. This means that in the leading order in $b$ at the points under consideration (on the singularity ring) there are no invariants measuring the curvature. Meanwhile, the vanishing of the Kretschmann scalar follows immediately upon the direct use of the found general form of $R_{\lambda \mu \nu \rho}$ (\ref{Rlmnr-on-ring}),
\begin{equation}
R_{\lambda \mu \nu \rho} R^{\lambda \mu \nu \rho} = \frac{r_g^2}{b^4} \left ( n^\lambda n^\mu \Delta_\lambda \Delta_\mu \Re \phi \right )^2 = 0 ,
\end{equation}

\noindent since $n^\lambda \Delta_\lambda$ gives zero when acting on the metric in the leading order in $b$, as mentioned above.

Some non-invariant characteristic of the Riemann tensor can be given, such as a mean value of its component. The positive contribution to $R_{\lambda \mu \nu \rho} R^{\lambda \mu \nu \rho}$ is due to $R_{k l m n}$ and $R_{0 k 0 l}$, 72 components in total (say $R_{k l m n}$ and $- R_{l k m n}$ are considered different components), the negative contribution is due to $R_{0 k l m}$, 72 components in total. These contributions are equal in absolute value to get zero $R_{\lambda \mu \nu \rho} R^{\lambda \mu \nu \rho}$,
\begin{equation}
R_{\lambda \mu \nu \rho} R^{\lambda \mu \nu \rho} = R_{k l m n} R^{k l m n} + 4 R_{0 k l m} R^{0 k l m} + 4 R_{0 k 0 l} R^{0 k 0 l} ,
\end{equation}

\noindent so the root mean square value of the physical component of $R_{\lambda \mu \nu \rho}$ is
\begin{eqnarray}
\overline{ | R_{\hat{\lambda } \hat{ \mu } \hat{ \nu } \hat{ \rho }} | } = \sqrt{\frac{1 }{ 72 } \left ( R_{k l m n} R^{k l m n} + 4 R_{0 k 0 l} R^{0 k 0 l} \right )} = \sqrt{ - \frac{1 }{ 72 } \cdot 4 R_{0 k l m} R^{0 k l m} } \nonumber \\ = \frac{ 4 \sqrt{ 2 \pi } }{3^{7 / 6} \Gamma ( 1 / 6 ) \mu^{5 / 6}} \frac{r_g }{ a^{4 / 3 } b^{5 / 3} } .
\end{eqnarray}

In the continuum theory, the Kretschmann scalar is equal to \cite{deFelice}
\begin{eqnarray}
R_{\lambda \mu \nu \rho} R^{\lambda \mu \nu \rho} & = & 12 r_g^2 \frac{ r^6 - 15 a^2 r^4 \cos^2 \theta + 15 a^4 r^2 \cos^4 \theta - a^6 \cos^6 \theta }{ \left ( r^2 + a^2 \cos^2 \theta \right )^6} \\ & & \left ( = \Re \frac{12 r_g^2}{\left [ ( \bx + i \by )^2 \right ]^3} \right ) . \nonumber
\end{eqnarray}

\begin{figure}[ht]
\centerline{\includegraphics[width=7.0cm]{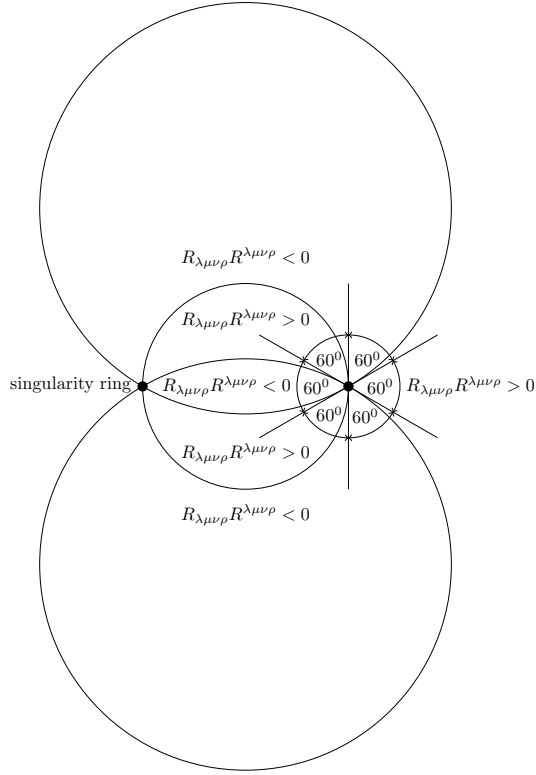}}
\caption{The sign of the continuum $R_{\lambda \mu \nu \rho} R^{\lambda \mu \nu \rho}$ in the space. \label{f2}}
\end{figure}

\noindent We can see that the space of $(x_1 , x_2 , x_3)$ is divided by three spheres into six regions, in each of which the $R_{\lambda \mu \nu \rho} R^{\lambda \mu \nu \rho}$ sign is constant. Each of these surfaces contains the singularity ring, and at any point of the ring, the tangent planes to them divide the space into six equal sectors of 60 degrees each, in which the signs of this value alternate, see Fig. \ref{f2}. Therefore, it is not surprising that the regularized value of $R_{\lambda \mu \nu \rho} R^{\lambda \mu \nu \rho}$ on the ring turns out to be zero.

\section{Source distribution}\label{sources}

The curvature includes a discrete analogue of the $\delta$-function with support at $x_3 = 0$ due to terms in $b^2 R_{\lambda \mu \nu \rho}$ of the type
\begin{equation}\label{nnd3d3-Re-phi}
n_\lambda n_\mu \bDelta_3 \Delta_3 \Re \phi \sim \delta_{k_3 0} ,
\end{equation}

\noindent $\delta_{k_3 0}$ is the Kronecker symbol arising from the double differentiation of the modulus $| k_3 |$ in the expansion of $\Re \phi$. This is also singled out by a larger value $O \left ( 1 \right )$ compared to the above $O \left ( b^{1 / 3} \right )$ in the analyticity region $x_3 \neq 0$. Then we will show that this contribution just leads to $R_{\lambda \mu} \neq 0$, that is, to the gravity sources.

In (\ref{nnd3d3-Re-phi}), we use a more precise Hermitian definition $- \bDelta_3 \Delta_3 $ for the discrete version of $\partial_3^2$, similar to that used in the discrete Laplacian (\ref{Laplace}), which can be important at the nonanalyticity points (in this case, using $\Delta_3^2$ instead of $- \bDelta_3 \Delta_3 $ results in $\delta_{k_3 1}$ instead of $\delta_{k_3 0}$).

The dependence on $| k_3 |$ is also present in $n_\lambda$, $\lambda = 1, 2$, but with a smaller coefficient, $\sim b^{2 / 3}$, and the appropriate terms in $b^2 R_{\lambda \mu \nu \rho}$ are suppressed by a power of $b$, despite the large value $\sim b^{- 1 / 3}$ of $\Re \phi$,
\begin{equation}
(\Re \phi ) \bDelta_3 \Delta_3 (n_\lambda n_\mu ) \sim b^{1 / 3} \delta_{k_3 0} , ~~~ \lambda , \mu \neq 3 .
\end{equation}

The symbol $\delta_{k_3 0}$ can also arise from the (single) differentiation of $n_3 \sim \sgn k_3$ over $k_3 $, but the corresponding terms in the $\Delta \Delta g$ part of $R_{\lambda \mu \nu \rho}$ are suppressed by a power of $b$,
\begin{equation}
\Delta_\lambda ( \Delta_3 n_3 ) n_\mu \Re \phi = ( \Delta_\lambda \Delta_3 n_3 ) n_\mu \Re \phi + \dots \sim b^{1 / 3} \delta_{k_3 0} .
\end{equation}

In the $( \Delta g )^2$ part of $R_{\lambda \mu \nu \rho}$, factors such as
\begin{equation}
\Delta_3 ( l_\lambda l_3 ) = \Delta_3 ( n_\lambda n_3 \Re \phi ) \sim \delta_{k_3 0}
\end{equation}

\noindent could presumably contain the contribution of interest to us, since their product could give $\left ( \delta_{k_3 0} \right )^2 = \delta_{k_3 0}$. However, such a product $\left [ \Delta_3 \left ( l_\lambda l_3 \right ) \right ] \left [ \Delta_3 \left ( l_\mu l_3 \right ) \right ]$ can only appear in $R_{\lambda \mu \nu \rho}$ from $C_{3 , \lambda 3 } C_{3 , \mu 3 }$, but $C_{3 , \lambda 3}$ does not contain $\Delta_3 \left ( l_\lambda l_3 \right )$. The latter is contained in $C_{\lambda , 3 3}$, and the potentially interesting term is $C_{\lambda , 3 3} C_{\lambda , \mu \nu}$, $\lambda , \mu , \nu \neq 3$. This includes terms with $\delta_{k_3 0}$, but they are suppressed by $b^{1 / 3}$ compared to the leading one, (\ref{nnd3d3-Re-phi}),
\begin{equation}\label{D(nnRe-phi)D(nnRe-phi)}
\left [ \Delta_3 ( n_\lambda n_3 \Re \phi ) \right ] \left [ \Delta_\mu ( n_\nu n_\rho \Re \phi ) \right ] \sim b^{1 / 3} \delta_{k_3 0} , ~~~ \lambda , \mu , \nu , \rho \neq 3 .
\end{equation}

As a result, the leading discrete $\delta$-function-like (Kronecker symbol) contribution to $R_{\lambda \mu \nu \rho}$ in the neighborhood of the singularity ring takes the form
\begin{equation}\label{R3l3m-on-ring}
R_{ 3 \lambda 3 \mu} =\frac{r_g}{2 b^2} n_\lambda n_\mu \bDelta_3 \Delta_3 \Re \phi = \frac{2}{\sqrt{3 \mu}} \frac{r_g}{a b^2 } n_\lambda n_\mu \delta_{k_3 0} .
\end{equation}

The Riemann tensor at $x_3 \neq 0$ (\ref{Rlmnr-on-ring}) and in the vicinity of $x_3 = 0$ (\ref{R3l3m-on-ring}) differ in the values and spatial distributions, $b^{- 5 / 3 }$ and $b^{ - 2 } \delta_{k_3 0}$. Note that the contribution (\ref{Rlmnr-on-ring}) also occurs at $x_3 = 0$ {\it in addition} to (\ref{R3l3m-on-ring}), since these contributions come from different terms in $\Re \phi$, bilinear in $\bk$ for (\ref{Rlmnr-on-ring}) and proportional to $| k_3 |$ for (\ref{R3l3m-on-ring}) ((\ref{Rlmnr-on-ring}) is simply suppressed at $x_3 = 0$ by $b^{1 / 3}$ in comparison with (\ref{R3l3m-on-ring})). Therefore, it is appropriate to use the same expression for $R_{\lambda \mu \nu \rho}$ for all $k_3$,
\begin{eqnarray}\label{Rlmnr+R3l3m-on-ring}
R_{\lambda \mu \nu \rho} & = & \frac{r_g}{ 2 b^2 } \left ( - n_\lambda n_\rho \bDelta_{ ( \mu } \Delta_{ \nu ) } \Re \phi - n_\mu n_\nu \bDelta_{ ( \lambda } \Delta_{ \rho ) } \Re \phi \right. \nonumber \\ & & \left. + n_\mu n_\rho \bDelta_{ ( \lambda } \Delta_{ \nu ) } \Re \phi + n_\lambda n_\nu \bDelta_{ ( \mu } \Delta_{ \rho ) } \Re \phi \right ) .
\end{eqnarray}

In addition, (\ref{Rlmnr+R3l3m-on-ring}) through (\ref{R3l3m-on-ring}) is an exhaustive $\delta$-function-like contribution to $R_{\lambda \mu \nu \rho}$ in the leading order over metric variations also at points located at a sufficiently large distance (\ref{dist>>a^1/3b^2/3}) from the ring. Indeed, this contribution follows when twice differentiating $\Re \phi $ in the terms $\bDelta_3 \Delta_3 ( n_\lambda n_\mu \Re \phi )$. Here we can also differentiate twice $n_\lambda$, $\lambda = 1, 2$ (or $n_\mu$). This operation gives $\delta_{k_3 0}$ at $x_1^2 + x_2^2 < a^2$, that is, $x_3 = 0$ and $\Re \phi = 0$ in factors. So, for our purposes, only $\Re \phi$ needs to be differentiated twice in $\bDelta_3 \Delta_3 ( n_\lambda n_\mu \Re \phi )$. $\delta_{k_3 0}$ also appears in $\Delta_3 n_3$ or $\bDelta_3 n_3$, that is, in the terms $\bDelta_3 \Delta_\lambda ( n_3 n_\mu \Re \phi )$ or $\Delta_3 \bDelta_\lambda ( n_3 n_\mu \Re \phi )$. But $\delta_{k_3 0}$ provides $\Re \phi = 0$ for $x_1^2 + x_2^2 < a^2$ and, therefore, also $\Delta_\lambda \Re \phi = 0$, $\lambda = 1, 2$. So, such terms do not contribute to $\delta_{k_3 0}$. In the $( \Delta g )^2$-part of $R_{\lambda \mu \nu \rho}$, as in the above consideration of the same question in the vicinity of the ring, $\delta_{k_3 0}$ could arise due to the terms (\ref{D(nnRe-phi)D(nnRe-phi)}) from $\Delta_3 n_3$ or $\bDelta_3 n_3$. Again, this provides $\Re \phi = 0$ among the factors, and this contribution disappears.

Thus, in the leading order over metric variations, either in the vicinity of the ring, or at macroscopic distances (\ref{dist>>a^1/3b^2/3}) from the ring, the source contribution in the curvature is
\begin{eqnarray}
\delta_{k_3 0} \mbox{ part of } R_{\lambda \mu \nu \rho} = \delta_{k_3 0} \mbox{ part of } \frac{r_g}{ 2 b^2 } \left ( \bDelta_3 \Delta_3 \Re \phi \right ) \left ( n_\lambda n_\nu \delta_\mu^3 \delta_\rho^3 + n_\mu n_\rho \delta_\lambda^3 \delta_\nu^3 \right. \nonumber \\ \left. - n_\lambda n_\rho \delta_\mu^3 \delta_\nu^3 - n_\mu n_\nu \delta_\lambda^3 \delta_\rho^3 \right ) .
\end{eqnarray}

\noindent Then we have for the Einstein tensor $G_{\lambda \mu } = R_{\lambda \mu } - R ( \eta_{\lambda \mu} + l_\lambda l_\mu ) / 2$,
\begin{eqnarray}
 & & \delta_{k_3 0} \mbox{ part of } G_{\lambda \mu } = \delta_{k_3 0} \mbox{ part of } \frac{r_g}{ 2 b^2 } \left ( \bDelta_3 \Delta_3 \Re \phi \right ) m_{\lambda \mu}, \nonumber \\ & & \hspace{-10mm} m_{\lambda \mu} = n_\lambda n_\mu + n_3^2 \eta_{ \lambda \mu } - n_3 n_\lambda \delta_\mu^3 - n_3 n_\mu \delta_\lambda^3 = \left ( \begin{array}{cccc} 1 - n_3^2 & n_1 & n_2 & 0 \\ n_1 & 1 - n_2^2 & n_1 n_2 & 0 \\ n_2 & n_1 n_2 & 1 - n_1^2 & 0 \\ 0 & 0 & 0 & 0 \end{array} \right ) .
\end{eqnarray}

\noindent On the left-hand side, we know a priori that the Kerr metric (reproduced at macroscopic distances (\ref{dist>>a^1/3b^2/3}) from the ring with our discrete $\phi$ (\ref{phi-cont}) in the leading order over metric variations) is a vacuum solution outside sources, and our discrete solution in the vicinity of the ring obeys $R_{\lambda \mu } = 0$ at $x_3 \neq 0$ (\ref{Rlm=0}), that is,
\begin{equation}
\delta_{k_3 0} \mbox{ part of } G_{\lambda \mu } = G_{\lambda \mu } .
\end{equation}

\noindent On the right-hand side, $\phi$ is an analytic continuation of the solution to the Poisson equation with a point source, and we can check that it is itself a vacuum solution outside $\delta$-function-like sources. Indeed, a simple calculation for the exact (within the framework of our ansatz, of course) $\phi$ (\ref{phi-discr}) gives
\begin{equation}\label{DDphi}
\sum_{j = 1}^3 \bDelta_j \Delta_j \phi ( \bx ) = \delta_{k_3 0} \int_{- \pi }^\pi \int_{- \pi }^\pi \frac{ \d q_1 \d q_2 }{ \pi b } \exp \left \{ i \frac{ s }{ b } \left [ q_1 x_1 + q_2 x_2 - a \kappa \right ] - \frac{ | x_3 | }{ b } \kappa \right \} .
\end{equation}

\noindent In passing, we note the following relation useful for estimating if $\phi$ is known:
\begin{equation}
\sum_{j = 1}^3 \bDelta_j \Delta_j \phi ( \bx ) = - 2 \delta_{k_3 0} \sh \left ( b \frac{\partial}{\partial | x_3 |} \right ) \phi .
\end{equation}

\noindent Thus,
\begin{equation}
\delta_{k_3 0} \mbox{ part of } \sum_{j = 1}^3 \bDelta_j \Delta_j \Re \phi = \sum_{j = 1}^3 \bDelta_j \Delta_j \Re \phi .
\end{equation}

\noindent Here, in turn,
\begin{equation}
\delta_{k_3 0} \mbox{ part of } \sum_{j = 1}^3 \bDelta_j \Delta_j \Re \phi = \delta_{k_3 0} \mbox{ part of } \bDelta_3 \Delta_3 \Re \phi ,
\end{equation}

\noindent since the differentiations over $k_1$, $k_2$ do not lead to $\delta$-function analogs. Thus,
\begin{equation}
G_{\lambda \mu } = \frac{r_g}{ 2 b^2 } \left ( \sum_{j = 1}^3 \bDelta_j \Delta_j \Re \phi \right ) m_{\lambda \mu} .
\end{equation}

\noindent The matrix structure of $G_{\lambda \mu }$ is such that
\begin{equation}\label{Gl=0}
G_{\lambda \mu } n^\mu = 0, \mbox{ that is, } G_{\lambda \mu } l^\mu = 0 .
\end{equation}

\noindent Due to this, the indices of $G_{\lambda \mu }$ can be raised/lowered using $\eta_{\lambda \mu}$.

As a consistency check, consider the vanishing of the divergence of $G_{\lambda \mu }$. For consistency, consider this at different distances, although this may be of particular interest for the vicinity of the singularity ring, while at sufficiently large distances from the ring, the formalism is close to continual, where such an identity is fulfilled in a routine way. Denoting the covariant finite difference by $D_\lambda$, we have
\begin{eqnarray}
D_\lambda G_\mu^\lambda & = & \frac{1}{ \sqrt{ - g }} \Delta_\lambda \left ( \sqrt{ - g } G_\mu^\lambda \right ) -\frac{1}{2} G^{\lambda \nu } \Delta_\mu g_{\lambda \nu} = \Delta_\lambda G_\mu^\lambda - \frac{1}{2} G^{\lambda \nu } \Delta_\mu \left ( l_\lambda l_\mu \right ) \nonumber \\ & = & \Delta_\lambda G_\mu^\lambda .
\end{eqnarray}

\noindent The neglect of the second term in the leading order over metric variations in the last equality is possible due to (\ref{Gl=0}).

At macroscopic distances from the ring, taking our $\phi$ (\ref{phi-cont}) or the known $l_0^2 / r_g$ (\ref{l0^2/r_g}), we find
\begin{eqnarray}\label{sum-b3-DD-phi-disk}
\sum_{j = 1}^3 \bDelta_j \Delta_j \Re \phi = - \frac{2 a b}{\left ( a^2 - x_1^2 - x_2^2 \right )^{3 / 2}} \delta_{k_3 0} , \\ G_{\lambda \mu } = -\frac{r_g }{b } \frac{ a }{\left ( a^2 - x_1^2 - x_2^2 \right )^{3 / 2}} \delta_{k_3 0} m_{\lambda \mu} ,
\end{eqnarray}

\noindent and the direct calculation gives the vanishing $D_\lambda G_\mu^\lambda$ in the leading order over metric variations,
\begin{eqnarray}
& & D_\lambda G_\mu^\lambda = \eta^{\lambda \nu} \Delta_\nu G_{\lambda \mu } \nonumber \\ & & = -\frac{r_g }{a b} \delta_{k_3 0} \left ( \begin{array}{rrrr} - \Delta_0 , & \Delta_1 , & \Delta_2 , & \Delta_3 \end{array} \right ) \left [ \left ( \begin{array}{cccc} x_1^2 + x_2^2 & - a x_2 & a x_1 & 0 \\ - a x_2 & a^2 - x_1^2 & - x_1 x_2 & 0 \\ a x_1 & - x_1 x_2 & a^2 - x_2^2 & 0 \\ 0 & 0 & 0 & 0 \end{array} \right ) \right. \nonumber \\ & & \left. \cdot \frac{1}{\left ( a^2 - x_1^2 - x_2^2 \right )^{3 / 2}} \right ] = \left ( \begin{array}{rrrr} 0 , & 0 , & 0 , & 0 \end{array} \right ) .
\end{eqnarray}

In the vicinity of the singularity ring, $n_3$ has smallness $O \left ( b^{1 / 3} \right )$, and in the leading order over metric variations $m_{\lambda \mu} = n_\lambda n_\mu$ (for $\lambda , \mu \neq 3$; otherwise $0$) and
\begin{eqnarray}\label{sum-b3-DD-phi-ring}
& & \sum_{j = 1}^3 \bDelta_j \Delta_j \Re \phi = \frac{4}{ \sqrt{3 \mu} } \frac{1}{a } \delta_{k_3 0} , \\ G_{\lambda \mu } & = & \frac{2}{ \sqrt{3 \mu} } \frac{r_g}{a b^2 } \delta_{k_3 0} n_\lambda n_\mu , \lambda , \mu \neq 3 , \mbox{ otherwise } G_{\lambda \mu } = 0 , \\ \label{DG-ring} \hspace{0mm} D_\lambda G_\mu^\lambda & = & \frac{ r_g }{ 2 b^2 } \Delta_\lambda m_\mu^\lambda \sum_{j = 1}^3 \bDelta_j \Delta_j \Re \phi = \frac{ r_g }{ 2 b^2 } n_\mu n^\alpha \Delta_\alpha \sum_{j = 1}^3 \bDelta_j \Delta_j \Re \phi = 0 ,
\end{eqnarray}

\noindent where $\alpha = 1 , 2$, $\mu \neq 3$ in (\ref{DG-ring}), but for $\mu = 3$ we get the same zero $D_\lambda G_\mu^\lambda$ identically. Here, since our expansion of metric functions goes in $b^{1 / 3} \bk$, the normal order of magnitude associated with $\Delta_\lambda$ is $O \left ( b^{1 / 3} \right )$ (this is obtained by the action of $\Delta_\lambda$ on $\sum_{j = 1}^3 \bDelta_j \Delta_j \Re \phi$; $O \left ( b^{2 / 3} \right )$ is obtained when $\Delta_\lambda$ acts on $n_\mu$, $\mu = 1, 2$). Now, in (\ref{DG-ring}), a combination $n_1 \Delta_1 + n_2 \Delta_2$ is formed, with which the order of magnitude $O \left ( b^{2 / 3} \right )$ is associated with the additional smallness $O \left ( b^{1 / 3} \right )$, as we considered above, (\ref{Delta-on-functions-orders}- \ref{nDelta-on-functions-orders}). (In the leading order, $n_1 \Delta_1 + n_2 \Delta_2 = - \sin \varphi_0 \Delta_1 + \cos \varphi_0 \Delta_2$ can be considered as a finite-difference version of the derivative over $\varphi$, so that the independence of $\sum_{j = 1}^3 \bDelta_j \Delta_j \Re \phi$ on $\varphi$ is essential here.) This just gives the vanishing $D_\lambda G_\mu^\lambda$ in the leading order over metric variations or, here, over $b$.

(\ref{DDphi}) implicitly uses $a >> b$ (via order of magnitude estimates, (\ref{nnd3d3-Re-phi}-\ref{D(nnRe-phi)D(nnRe-phi)})), but for comparison it is useful to formally consider its right-hand side at $a = 0$,
\begin{eqnarray}
\sum_{j = 1}^3 \bDelta_j \Delta_j \phi ( \bx ) & = & \frac{ \delta_{k_3 0} }{ \pi b } \left [ \int_{ - \pi }^\pi \exp \left ( i q_1 \frac{x_1}{b} \right ) \d q_1 \right ] \left [ \int_{ - \pi }^\pi \exp \left ( i q_2 \frac{x_2}{b} \right ) \d q_2 \right ] \nonumber \\ & = & \frac{ 4 \pi }{ b } \delta_{ m_1 0 }  \delta_{ m_2 0 } \delta_{k_3 0} , ~~~ x_j = m_j b , ~~ j = 1, 2 .
\end{eqnarray}

\noindent This is an analogue of the continuum $- \bnabla^2 \phi = 4 \pi \delta^3 ( \bx )$.

At $a \neq 0$, consider an analogue of the continuum $\int \left ( - \bnabla^2 \phi \right ) \d^3 \bx$, that is, the sum
\begin{eqnarray}
& & \sum_{m_1 m_2 k_3} b^3 \sum_{j = 1}^3 \frac{1}{b^2 } \bDelta_j \Delta_j \phi = \sum_{m_1 = - N}^N \sum_{m_2 = - N}^N \int_{ - \pi }^\pi \int_{ - \pi }^\pi \frac{\d^2 q}{ \pi } \exp \left \{ i s \left [ q_1 m_1 + q_2 m_2 \vphantom{\frac{a}{b}} \right. \right. \nonumber \\ & & \left. \left. - \frac{a}{b} \kappa \left ( q_1 , q_2 \right ) \right ] \right \} = \int_{ - \pi }^\pi \int_{ - \pi }^\pi \frac{\d^2 q}{ \pi } \frac{ \sin [ q_1 ( N + 1 / 2 )]}{\sin (q_1 / 2)} \frac{ \sin [ q_2 ( N + 1 / 2 )]}{\sin (q_2 / 2)} \nonumber \\ & & \cdot \exp \left [ - i \frac{a}{b} \kappa \left ( q_1 , q_2 \right ) \right ] .
\end{eqnarray}

\noindent Here $N >> 1$. At $N >> a / b$, the sines effectively behave as $\delta$-functions here,
\begin{equation}
\frac{ \sin [ q_1 ( N + 1 / 2 )]}{\sin (q_1 / 2)} \Rightarrow 2 \pi \delta ( q ) ,
\end{equation}

\noindent and we have
\begin{equation}\label{sum-b3-DD-phi-full}
\sum_{m_1 m_2 k_3} b^3 \sum_{j = 1}^3 \frac{1}{b^2 } \bDelta_j \Delta_j \phi = 4 \pi ,
\end{equation}

\noindent the same, as in the case $a = 0$.

But at $a \neq 0$, the source is not pointlike. The contribution of the inner vertices of the disk $x_1^2 + x_2^2 < a^2$, located at macroscopic distances (\ref{dist>>a^1/3b^2/3}) from the ring, to the sum (\ref{sum-b3-DD-phi-full}) can be estimated by passing to an integral of the discrete Laplacian of $\phi$ (\ref{sum-b3-DD-phi-disk}) with the lower cutoff $\sim a^{1 / 3} b^{2 / 3}$ for these distances,
\begin{equation}\label{sum-disk}
\sum_{\rm disk} b^3 \sum_{j = 1}^3 \frac{1}{b^2 } \bDelta_j \Delta_j \phi = - \int \frac{2 a \delta ( x_3 ) }{\left ( a^2 - x_1^2 - x_2^2 \right )^{3 / 2}} \d^3 \bx \sim - \sqrt[3]{\frac{a }{b }} .
\end{equation}

\noindent This contribution is infinite at $b \to 0$ and has the sign opposite to that required.

The contribution to this sum from the neighborhood of the ring should also be taken into account. To estimate, we take the value of the discrete Laplacian of $\phi$ to be (\ref{sum-b3-DD-phi-ring}) in a band along the ring with width $\sim a^{1 / 3} b^{2 / 3}$. The number of vertices in this band is the area of the band divided by $b^2$, and it is $\sim (a / b)^{4 / 3}$. The corresponding contribution to the sum (\ref{sum-b3-DD-phi-full}) is
\begin{equation}
\sum_{\rm ring} b^3 \sum_{j = 1}^3 \frac{1}{b^2 } \bDelta_j \Delta_j \phi \sim \frac{b }{a } \left ( \frac{a }{b } \right )^{4 / 3}  \sim \sqrt[3]{\frac{a }{b }} .
\end{equation}

\noindent This value has the same degree of divergence at $b \to 0$ as (\ref{sum-disk}), but has the required sign. The sum (\ref{sum-b3-DD-phi-full}) is the result of the mutual cancellation of large numbers.

In the continuum GR, we have
\begin{eqnarray}\label{dd-phi-cont}
-\bnabla^2 \phi & = & \delta ( x_3 ) \iint \frac{\d^2 p}{ \pi } \exp \left \{ i \left [ p_1 x_1 + p_2 x_2 - ( a - i0 ) \sqrt{p_1^2 + p_2^2} \right ] \right \} \nonumber \\ & = & -2 \lim_{\varepsilon \to 0} \frac{(a - i \varepsilon) \delta ( x_3 )}{ \left [ (a - i \varepsilon)^2 - x_1^2 - x_2^2 \right ]^{3 / 2} } .
\end{eqnarray}

\begin{figure}[ht]
\centerline{\includegraphics[width=7.0cm]{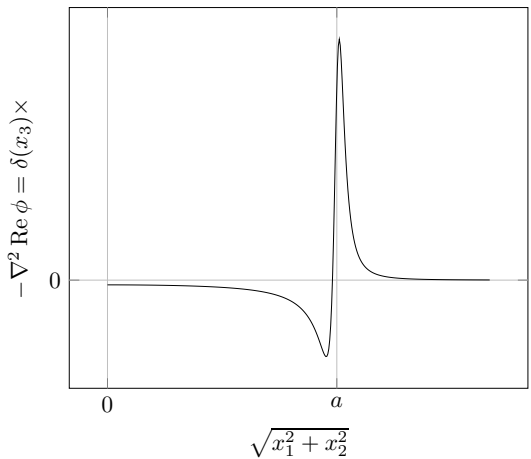}}
\caption{The source distribution, regularized in the continuum (\ref{dd-phi-cont}) (or qualitative in the discrete approach; the factor at $\delta ( x_3 )$, as written here, or at $\delta_{k_3 0}$, respectively). \label{f3}}
\end{figure}

\noindent For a small but nonzero $\varepsilon$, the multiplier at $\delta ( x_3 )$ here depends on $x_1$, $x_2$ (i.e., on $\sqrt{x_1^2+x_2^2}$), as shown in Fig. \ref{f3}.
Trying to write the limiting $\varepsilon \to 0$ distribution in closed form as an ordinary function, we get something like
\begin{equation}
-\bnabla^2 \Re \phi = - \frac{ 2 a \theta \left ( a^2 - x_1^2 - x_2^2 \right )}{\left ( a^2 - x_1^2 - x_2^2 \right )^{3 / 2}} \delta ( x_3 ) + \frac{1 }{ 0 } \delta \left ( a - \sqrt{ x_1^2 + x_2^2 } \right ) \delta ( x_3 ) .
\end{equation}

\noindent The $\delta$-function term with support on the ring turns out to have a formally infinite coefficient. In the above discrete approach, such a term is finite and phenomenologically in the continuum notations looks as
\begin{equation}
-\bnabla^2 \Re \phi \sim \dots + a^{- 2 / 3} b^{- 1 / 3} \delta \left ( a - \sqrt{ x_1^2 + x_2^2 } \right ) \delta ( x_3 ) .
\end{equation}

\section{Conclusion}

In the functional integral framework, the problem of the discrete Kerr geometry (like the problem of any other discrete classical solution) arises as the problem of finding the optimal background metric for the perturbative expansion within this framework. In addition to the classical equations of motion (\ref{dS/dl=0}), we also have the condition of maximizing the functional measure (\ref{def-l0}) (once the measure turns out to have maxima), which fixes the elementary length scale $b$. In the leading order over metric variations from simplex to simplex, the equations of motion are reduced to the finite-difference form of Einstein's equations.

To solve analytically in general form the full Einstein system, especially discrete, seems to be an unattainable task, but we can use an ansatz with one unknown complex function $\phi$ which gives the Kerr metric in the continuum case. In the discrete case, we obtain a solution close to the continuum Kerr one at large distances, having a finite metric and effective curvature on the ring that was singular in the continuum.

A distinctive feature of this system is that the typical distance from the singularity at which the metric approaches the continuum one is $\sqrt[3]{a b^2}$ which is much larger than such a distance $b$ in the discrete Schwarzschild case (we assume $a >> b$, since this case is physically more significant). That is, the lattice effects in the vicinity of the singularity are enhanced here in comparison with the Schwarzschild case.

The Riemann tensor components are found in the vicinity of the ring. Those with sources $R_{\lambda \mu} \neq 0$ have support in the plane $x_3 = 0$ on a disk which is a $O ( \sqrt[3]{a b^2} ) $-neighborhood of the disk $x_1^2 + x_2^2 \leq a^2$ (formed by the singularity ring). The vacuum $ R_{\lambda \mu \nu \rho} $ (i.e., at $x_3 \neq 0$) in the vicinity of the ring is $O ( b^{ - 5 / 3 } ) $ in the leading order at small $b$. It has $ R_{\lambda \mu \nu \rho} R^{\lambda \mu \nu \rho} = 0 $, and all four curvature invariants vanish on the ring in this order. In the neighborhood of the singularity ring at $x_3 = 0$, $ R_{\lambda \mu \nu \rho} $ has $R_{\lambda \mu} \neq 0$ and is $O ( b^{-2} )$.

The source distribution $G_{\lambda \mu}$ in the continuum can be written as a limit of a regularized expression (or a derivative of a generalized function). Heuristically, it can be represented as the sum of the $\delta$-function on the ring with an {\it infinite} coefficient and a smooth function of the opposite sign on the disk formed by the ring, diverging when approaching the ring. In the discrete version, there are no infinities; phenomenologically, $G_{\lambda \mu}$ can be written as the sum of the $\delta$-function on the ring with a coefficient $ O ( a^{- 2 / 3} b^{- 1 / 3} ) $ and a smooth function of the opposite sign on the disk, cut off at a distance $O ( \sqrt[3]{a b^2} ) $ from the ring. Integrally, the large (at small $b$) contribution of the ring is partially cancelled by the inner part of the disk, and this corresponds to $r_g$, as in the Schwarzschild problem.

Physically, we have a rather exotic required energy-momentum tensor, such that the total energy on the ring is many times (for small $b$) greater than the mass that determines $r_g$, and this excess is in total compensated by the negative contribution of the inner part of the disk.

The fulfillment of the discrete version of the conservation law $D_\lambda G_\mu^\lambda = 0$ in the leading order over metric variations is also verified, including, which is especially interesting, on the singularity ring.

\section*{Acknowledgments}

The present work was supported by the Ministry of Education and Science of the Russian Federation.


\begin{thebibliography}{99}
\bibitem{Regge}
 T. Regge, General relativity theory without coordinates, {\it Nuovo Cimento} {\bf 19}, 558 (1961).
\bibitem{Fein}
 G. Feinberg, R. Friedberg, T. D. Lee, and M. C. Ren, Lattice gravity near the continuum
 limit, {\it Nucl. Phys. B} {\bf 245}, 343 (1984).
\bibitem{CMS}
 J. Cheeger, W. M\"{u}ller, and R. Shrader, On the curvature of the piecewise flat spaces,
 {\it Commun. Math. Phys.} {\bf 92}, 405 (1984).
\bibitem{Ham}
 H. W. Hamber, Quantum Gravity on the Lattice, {\it Gen. Rel. Grav.} {\bf 41}, 817 (2009); arXiv:0901.0964[gr-qc].
\bibitem{HamWil1}
 H. W. Hamber and R. M. Williams, Newtonian Potential in Quantum Regge Gravity, {\it Nucl.Phys. B} {\bf 435}, 361 (1995); ({\it Preprint} arXiv:hep-th/9406163).
\bibitem{HamWil2}
 H. W. Hamber and R. M. Williams, On the Measure in Simplicial Gravity, {\it Phys. Rev. D} {\bf 59},  064014 (1999); ({\it Preprint} arXiv:hep-th/9708019).
\bibitem{cdt}
 J. Ambjorn, A. Goerlich, J. Jurkiewicz, and R. Loll,  Nonperturbative Quantum Gravity, {\it Physics Reports} {\bf 519}, 127 (2012); arXiv:1203.3591[hep-th].
\bibitem{Mik}
 A. Mikovi\'{c} and M. Vojinovi\'{c}, Quantum gravity for piecewise flat spacetimes, Proceedings of the MPHYS9 conference, 2018; 	arXiv:1804.02560[gr-qc].
\bibitem{WilCol}
 P. A. Collins and R. M. Williams, Dynamics of the Friedmann universe using Regge calculus,  {\it Phys. Rev. D} {\bf 7}, 965 (1973).
\bibitem{Gen}
 A. P. Gentle, A cosmological solution of Regge calculus, {\it Class. Quantum Grav.} {\bf 30}, 085004 (2013); arXiv:1208.1502[gr-qc].
\bibitem{Bre4}
 L C Brewin, A numerical study of the Regge calculus and Smooth Lattice methods on a Kasner cosmology, {\it Classical and Quantum Gravity} {\bf 32}, 195008 (2015); arXiv:1505.00067[gr-qc].
\bibitem{WilLiu}
 R. G. Liu and R. M. Williams, Regge calculus models of closed lattice universes, {\it Phys. Rev. D} {\bf 93}, 023502 (2016); arXiv:1502.03000[gr-qc].
\bibitem{GlaLol}
 L. Glaser and R. Loll, CDT and cosmology, {\it Comptes Rendus Physique} {\bf 18}, 265 (2017); arXiv:1703.08160[gr-qc].
\bibitem{Wong}
 C.-Y. Wong, Application of Regge calculus to the Schwarzshild and Reissner-Nordstrøm geometries, {\it Journ. Math. Phys.} {\bf 12}, 70 (1971).
\bibitem{Bre2}
 L. Brewin, Einstein-Bianchi system for smooth lattice general relativity. I. The Schwarzschild spacetime, {\it Phys. Rev. D} {\bf 85}, 124045 (2012); arXiv:1101.3171[gr-qc].
 \bibitem{Ash1}
 A. Ashtekar, J. Olmedo and P. Singh, Quantum  Transfiguration  of Kruskal Black Holes, {\it Phys. Rev. Lett.} {\bf 121}, 241301 (2018); arXiv:1806.00648[gr-qc].
\bibitem{Ash2}
 A. Ashtekar, J. Olmedo and P. Singh, Quantum  extension  of  the Kruskal spacetime, {\it Phys. Rev. D} {\bf 98}, 126003 (2018); arXiv:1806.02406[gr-qc].
\bibitem{Kha1}
 V. M. Khatsymovsky, On the discrete version of the black hole solution, {\it Int. J. Mod. Phys. A} {\bf 35}, 2050058 (2020); arXiv:1912.12626[gr-qc].
\bibitem{our}
 V. M. Khatsymovsky, On the discrete version of the Schwarzschild problem, {\it Universe} {\bf 6}, 185 (2020);  arXiv:2008.13756[gr-qc].
\bibitem{our1}
 V. M. Khatsymovsky, On the non-perturbative graviton propagator, {\it Int. J. Mod. Phys. A} {\bf 33}, 1850220 (2018); arXiv:1804.11212[gr-qc].
\bibitem{Fro}
 J. Fr\"{o}hlich, Regge calculus and discretized gravitational functional integrals, in {\it Nonperturbative Quantum Field Theory: Mathematical Aspects and Applications, Selected Papers} (World Scientific, Singapore, 1992), p. 523, IHES preprint 1981 (unpublished).
\bibitem{Kha}
 V. M. Khatsymovsky, Tetrad and self-dual formulations of Regge calculus, {\it Class. Quantum Grav.} {\bf 6}, L249 (1989).
\bibitem{ADM1}
 R. Arnowitt, S. Deser, and C. W. Misner, The Dynamics of General Relativity, in {\it Gravitation: an introduction to current research, Louis Witten ed.} (Wiley, 1962), chapter 7, p. 227; arXiv:gr-qc/0405109[gr-qc].
\bibitem{our2}
 V. M. Khatsymovsky, On the discrete Christoffel symbols, {\it Int. J. Mod. Phys. A} {\bf 34}, 1950186 (2019); arXiv:1906.11805[gr-qc].
\bibitem{RocWil}
 M. Rocek and R. M. Williams, The quantization of Regge calculus, {\it Z. Phys. C} {\bf 21}, 371 (1984).
\bibitem{Baines}
 J. Baines, T. Berry, A. Simpson and M. Visser, Unit-lapse versions of the Kerr spacetime, {\it Class. Quantum Grav.} {\bf 38}, 055001 (2021); arXiv:2008.03817 [gr-qc].
\bibitem{Boyer}
 R. H. Boyer and R. W. Lindquist, Maximal Analytic Extension of the Kerr Metric, {\it Journ. Math. Phys.} {\bf 8} 265 (1967).
\bibitem{Kha2}
 V. M. Khatsymovsky, On the Kerr metric in a synchronous reference frame, arXiv:2101.07147 [gr-qc], submitted to {\it Int. J. Mod. Phys. D}.
\bibitem{Doran}
 C. Doran, New form of the Kerr solution, {\it Phys. Rev. D} {\bf 61}, 067503 (2000); arXiv:gr-qc/9910099.
\bibitem{Kerr}
 R. P. Kerr, Gravitational field of a spinning mass as an example of algebraically special metrics, {\it Phys. Rev. Lett.} {\bf 11}, 237-238 (1963).
\bibitem{Kerr1}
 R. P. Kerr, Gravitational collapse and rotation, published in: {\it Quasi-stellar sources and gravitational collapse:  Including the proceedings of the First Texas Symposium on Relativistic Astrophysics, I. Robinson, A. Schild, and E. L. Schucking eds.} (University of Chicago Press, Chicago, 1965), pages 99–103.
\bibitem{Chandra}
 S. Chandrasekhar, {\it The mathematical theory of black holes} (Oxford, Clarendon Press. 1983).
\bibitem{Newman}
 E. T. Newman and A. I. Janis, Note on the Kerr Spinning-Particle Metric, {\it J. Math. Phys.} {\bf 6}, 915 (1965).
\bibitem{Rajan}
 D. Rajan and M. Visser, Cartesian Kerr–Schild variation on the Newman–Janis trick, {\it Int. J. Mod. Phys. D} {\bf 26}, 1750167 (2017); arXiv:1601.03532[gr-qc].
\bibitem{deFelice}
 F. de Felice and M. Bradely, Rotational anisotropy and repulsive effects in the Kerr metric, {\it Class. Quantum Grav.} {\bf 5}, 1577 (1988).

\end{thebibliography}
\end{document}